\documentclass[a4paper,11pt]{article}
\pdfoutput=1
\usepackage{jheppub} 
\usepackage[T1]{fontenc}
\usepackage{slashed}
\usepackage{braket}
\usepackage{multirow,arydshln,rotating}
\usepackage{afterpage}
\usepackage{multirow}
\usepackage{arydshln}

\def\ba{\begin{align}}
\def\pa{\partial}

\def\nn{\nonumber}
\def\be{\begin{equation}}
\def\ee{\end{equation}}
\def\bea{\begin{eqnarray}}
\def\eea{\end{eqnarray}}
\def\a{\alpha}
\def\b{\beta}
\def\g{\gamma}
\def\d{\delta}
\def\D{\Delta}
\def\l{\lambda}

\def\f{\phi}

\def\nn{\nonumber}

\def\le{\left}
\def\ri{\right}
\def\cO{{\cal O}}

\def\e{\epsilon}

\def\m{\mu}
\def\n{\nu}

\def\r{\rho}
\def\s{\sigma}

\def\z{\zeta}
\def\sq
\def\y{\psi}

\def\dif{\mathrm{d}}

\title{Second-Order Hydrodynamics and Universality in Non-Conformal Holographic Fluids}
\author{Philipp Kleinert and}
\author{Jonas Probst}
\affiliation{Rudolf Peierls Centre for Theoretical Physics, University of Oxford, \\
1 Keble Road, Oxford OX1 3NP, United Kingdom}
\emailAdd{philipp.kleinert@physics.ox.ac.uk}
\emailAdd{jonas.probst@physics.ox.ac.uk}

\abstract{We study second-order hydrodynamic transport in strongly coupled non-con\-for\-mal field theories with holographic gravity duals in asymptotically anti-de Sitter space. We first derive new Kubo formulae for five second-order transport coefficients in non-conformal fluids in $(3+1)$ dimensions. We then apply them to holographic RG flows induced by scalar operators of dimension $\D=3$. For general background solutions of the dual bulk geometry, we find explicit expressions for the five transport coefficients at infinite coupling and show that a specific combination, $\tilde{H}=2\eta\tau_\pi-2\le(\kappa-\kappa^*\ri)-\l_2$, always vanishes. We prove analytically that the Haack-Yarom identity $H=2\eta\tau_\pi-4\l_1-\l_2=0$, which is known to be true for conformal holographic fluids at infinite coupling, also holds when taking into account leading non-conformal corrections. The numerical results we obtain for two specific families of RG flows suggest that $H$ vanishes regardless of conformal symmetry. Our work provides further evidence that the Haack-Yarom identity $H=0$ may be universally satisfied by strongly coupled fluids.}

\preprint{OUTP-16-22P}

\begin{document} 
\maketitle
\flushbottom

\section{Introduction and summary}

Hydrodynamics \cite{LandauLifshitz,Kovtun:2012rj} is the low-energy effective theory for slowly varying fluctuations around thermal equilibrium. All relevant dynamics in this regime is captured by the conservation laws of the underlying microscopic field theory. In the simplest case of an uncharged relativistic fluid in $(3+1)$ dimensions, energy and momentum density are the only four conserved charges. These charges themselves or alternatively their respective conjugate variables, temperature and fluid velocity, are the only relevant degrees of freedom and constitute the fluid variables. The corresponding four equations of motion are given by the conservation of the stress-energy tensor, supplemented by constitutive relations that express the stress-energy tensor in terms of the four fluid variables. The constitutive relations take the form of a systematic expansion in gradients of the fluid variables, gradients which are assumed to be small in the hydrodynamic regime. In the spirit of effective field theory, all independent terms compatible with the underlying symmetries appear in the constitutive relation at a given order in gradients, each multiplied by a free parameter. These parameters are referred to as transport coefficients and provide the low-energy constants of hydrodynamics. In order to determine their values we need to compute appropriate real-time correlators of the stress tensor, both in the underlying microscopic theory and from hydrodynamics, and then match the two results. The constitutive relation for the stress tensor of an uncharged relativistic fluid features two transport coefficients at first order in gradients, the shear viscosity $\eta$ and the bulk viscosity $\z$, and another fifteen coefficients at second order \cite{Baier:2007ix,Romatschke:2009kr}. Kubo formulae, which tell us which correlators exactly to look at in order to compute a specific transport coefficient \cite{Kadanoff,Kovtun:2012rj}, are known for $\eta$ and $\z$, for the five second-order coefficients already present in conformal fluids \cite{Moore:2010bu}, and for six of the ten second-order coefficients only present in non-conformal fluids \cite{Moore:2012tc}.

In recent years, hydrodynamics has been successfully applied to describe the early-stage evolution of the quark-gluon plasma (QGP) created in heavy-ion collisions at RHIC \cite{Gyulassy:2004zy,Arsene:2004fa,Back:2004je,Adcox:2004mh,Adams:2005dq}. In fact, it captures the evolution of the QGP from surprisingly early times onwards. However, it is not sufficient to simulate the QGP with first-order viscous hydrodynamics, as this is plagued by superluminal modes which violate causality. To get rid of these modes it is necessary to take into account second-order terms as well \cite{Baier:2007ix}. As the temperature of the QGP is not too far from the confinement scale of QCD, the QGP falls into the intermediate-coupling regime and its transport coefficients cannot be computed by perturbative methods. Lattice calculations are not well-suited to computations of transport coefficients either: computations of real-time correlators face the formidable problem of analytic continuation whereas indirect methods are plagued with uncertainties \cite{Moore:2008ws}.

The only currently available tool that allows for the computation of real-time correlators in strongly interacting field theories is gauge/gravity duality \cite{Maldacena:1997re,Aharony:1999ti,Gubser:1998bc,Witten:1998qj,Son:2002sd,Herzog:2002pc}. In particular, it reduces computing stress-tensor correlators in the hydrodynamic regime to solving classical Einstein's equations on anti-de Sitter (AdS) backgrounds in a small-momentum expansion. Applying the appropriate Kubo formulae, one can read off the transport coefficients in the strongly coupled holographic field theory \cite{Policastro:2002se,Policastro:2002tn,Son:2007vk}. Apart from its relevance for strongly coupled field theories, a hydrodynamic interpretation of solutions to Einstein's equations is interesting in its own right, because it extends the analogy between thermodynamics and black-hole mechanics beyond thermal equilibrium to a more general correspondence between fluids and gravity \cite{Bhattacharyya:2008jc}.

At present, the construction of the exact holographic dual of realistic theories such as QCD is beyond reach. The best one can hope for when trying to make connections with experiment is to identify and investigate properties that hold for a large class of holographic models: being insensitive to the microscopic details of the dual field theory, it is possible that such properties are shared by many or even all QFTs in the strong coupling limit, including the ones realised in nature. The most prominent example of such a universal property is the ratio of shear viscosity $\eta$ over entropy density $s$ \cite{Policastro:2001yc} which is known to obey $\eta/s=1/4\pi$ for any strongly coupled theory with a two-derivative gravity dual and unbroken SO(3) rotational symmetry \cite{Kovtun:2003wp,Buchel:2003tz,Kovtun:2004de,Buchel:2004qq,Starinets:2008fb,Son:2007vk,Brustein:2008cg,Iqbal:2008by,Cremonini:2011iq}. Current estimates for the value of $\eta/s$ in the QGP extracted from heavy ion collision data are indeed close to $1/4\pi$ \cite{Teaney:2003kp,Shuryak:2003xe,Romatschke:2007mq,Song:2007fn,Dusling:2007gi,Song:2007ux,Luzum:2008cw}. 

Universality in hydrodynamic transport is most likely to be observed among transport coefficients that can be measured without considering sound perturbations, which would necessarily excite the model-specific matter content \cite{Benincasa:2005iv}. There exists one particular relation between second-order transport coefficients which promises to exhibit such universal behaviour:
\begin{align}
	H\equiv 2\eta\tau_\pi-4\l_1-\l_2=0\label{universalIdentity}\;.
\end{align}
Prompted by the observation in ref.~\cite{Erdmenger:2008rm}, Haack and Yarom \cite{Haack:2008xx} demonstrated that the combination $H$ vanishes for conformal holographic theories with a two-derivative gravity dual, and with any number of $U(1)$-charges at finite density. It has further been shown that $H$ remains zero when taking into account leading corrections to the infinite coupling limit, both in planar $\mathcal{N}=4$ \cite{Grozdanov:2014kva} and in the hypothetical fluid dual to Gauss-Bonnet gravity \cite{Shaverin:2012kv,Grozdanov:2015asa,Shaverin:2015vda}. A simple example of non-conformal holographic transport was studied in ref.~\cite{Bigazzi:2010ku}:  employing the method developed in ref.~\cite{Kanitscheider:2009as} and working to linear order in $1/3-c_s^2$, the authors showed that $H$ also vanishes for the non-conformal Chamblin-Reall background. This background, however, can simply be viewed as the analytic continuation of higher-dimensional AdS compactified on a torus \cite{Gubser:2008ny}. While the transport coefficients in this special non-conformal model do differ from their conformal values, they are nonetheless completely dictated by the higher-dimensional conformal theory \cite{Kanitscheider:2009as}. In particular, the five coefficients that are already present in the conformal case are simply multiplied by a common factor so that relations that hold between coefficients in the conformal case, such as $H=0$, trivially apply to this non-conformal compactification as well. The only holographic model in which non-conformal corrections to $H$ have been taken into account beyond leading order are the compactified D4-branes of ref.~\cite{Wu:2016erb} where the relation $H=0$ was also found to hold.\footnote{Other recent holographic and non-holographic studies of second-order transport in non-conformal relativistic fluids include refs.~\cite{Denicol:2012cn,Molnar:2013lta,Denicol:2014vaa,Denicol:2014mca,Jaiswal:2014isa,Finazzo:2014cna,Becattini:2015nva}.} Note that neither the Chamblin-Reall background studied in ref.~\cite{Bigazzi:2010ku} nor the compactified branes of ref.~\cite{Wu:2016erb} admit an asymptotically AdS region.\footnote{In both cases the bulk geometry can be viewed as a compactification of AdS and one can essentially borrow the higher dimensional AdS/CFT dictionary \cite{Mateos:2007vn,Kanitscheider:2008kd}.} Their holographic duals therefore do not have an obvious UV fixed point.

Whether $H=0$ holds more generally in holographic theories without conformal symmetry remains an open question which we want to address in this paper. To this end we compute the second-order transport coefficients entering $H$ for a large class of non-conformal holographic models from three-point functions of the stress tensor. The precise type of models we consider are holographic renormalisation-group (RG) flows in asymptotically $AdS_5$, triggered by an arbitrary scalar operator of dimension $\D=3$. We prove analytically that $H$ vanishes for this class of models even when leading non-conformal corrections to the transport coefficients are included.  We subsequently study two specific families of RG flows from that class and show numerically that $H$ vanishes along both.

The content of this paper is structured as follows: in section \ref{hydroSection} we derive new Kubo formulae for the five second-order coefficients
\begin{align}
	\kappa\;,\;\;\;\,\eta\,\tau_\pi+\kappa^*\;,\;\;\;\l_1+\frac{\kappa^*}{2}\;,\;\;\;\l_2\;,\;\;\;\l_3-2\kappa^*\;, \label{coeffs}
\end{align}
by considering the response of the stress tensor to shear perturbations of the external metric. The coefficients \eqref{coeffs} are combinations of the five coefficients present in conformal fluids and the non-conformal coefficient $\kappa^*$. The combination $H$, eq.~\eqref{universalIdentity}, can be obtained as a linear combination of these coefficients. The Kubo formulae we derive are valid for any uncharged relativistic fluid in four dimensions, with or without conformal symmetry. We end section \ref{hydroSection} with a brief explanation of how they can be applied specifically to theories with a holographic dual. In section \ref{model} we introduce the class of strongly coupled non-conformal models studied in this paper. These are four-dimensional holographic theories with a UV fixed point, deformed by a relevant scalar operator of dimension $\D=3$. The dual description of such RG flows is provided by Einstein gravity in asymptotically $AdS_5$, coupled to a scalar field with mass $m^2L^2=\D\le(\D-4\ri)=-3$ but otherwise arbitrary potential. We derive the relevant bulk equations of motion for black-brane backgrounds and for metric perturbations. In section \ref{einstein} we solve Einstein's equations for bulk metric perturbations around a general black-brane background. Section \ref{StressTensor} contains our analytical results on second-order transport in the class of non-conformal holographic models that we investigate. We present formulae for the five second-order coefficients \eqref{coeffs} for a given background solution in subsection \ref{FormulaeCoefficients}. The coefficients always satisfy the relation $\tilde{H}\equiv2\eta\tau_\pi-2\le(\kappa-\kappa^*\ri)-\l_2=0$. In subsection \ref{proof} we prove analytically that $H$ remains zero when leading non-conformal corrections to the transport coefficients are taken into account. Section \ref{num} contains our numerical results. Treating the bulk scalar field as a small perturbation, we obtain in subsection \ref{coeffsPerturb} the leading non-conformal corrections to the transport coefficients in the vicinity of the UV fixed point. These leading corrections only depend on the mass term in the scalar potential and therefore apply to all holographic RG flows triggered by a scalar operator of dimension $\D=3$. In subsection \ref{numFlows} we introduce two specific families of bulk potentials: the first one describes RG flows to a fixed point in the IR, the second one describes flows to a non-conformal IR. We present our numerical results for the five transport coefficients in subsection \ref{plots}. Our main result is that the combination $H$, eq.~\eqref{universalIdentity}, vanishes for both families of flows, even if the individual transport coefficients deviate from their conformal values by factors of two and more. In subsection \ref{entropyConstraints} we employ relations between second-order coefficients and their derivatives which hold if one demands that the local entropy production be positive under all circumstances \cite{Romatschke:2009kr,Bhattacharyya:2012nq}\footnote{Ref.~\cite{Jensen:2012jh} obtains the same relations by coupling the fluid to external sources.}. This allows us to extend our results to eight second-order coefficients. We conclude with a summary of our results and propose future directions of research in section \ref{conclusion}. We attach technical details of our calculations in appendices \ref{AppendixConstituent}--\ref{ChamblinReall}.


\section{New Kubo formulae for non-conformal second-order hydrodynamics}\label{hydroSection}

We begin this section with a brief introduction to hydrodynamics in subsection \ref{recap}. We then derive a new set of Kubo formulae \eqref{Kubo} for five second-order transport coefficients in subsection \ref{hydroResponse}. These Kubo formulae are valid for any uncharged relativistic fluid in $(3+1)$ dimensions, with or without conformal symmetry, and constitute one of the main results of this paper. In subsection \ref{HologComput} we outline how they can be applied specifically to strongly coupled fluids with a holographic gravity dual.

\subsection{Quick recapitulation of hydrodynamics}\label{recap}

Hydrodynamics \cite{LandauLifshitz,Kovtun:2012rj} is the effective low-energy description of slowly varying fluctuations around thermal equilibrium in an interacting field theory. It assumes that the only relevant degrees of freedom are the expectation values of global charge densities in small patches of the fluid that have already equilibrated. These patches of local equilibrium are taken to be large compared to the microscopic scales of the underlying field theory but approximately local compared to macroscopic thermodynamic scales. 

The fluid equations of motion are the conservation equations for the corresponding current densities. They must be supplemented by constitutive relations which express the current densities in terms of the charge densities. By definition, hydro fluctuations have small momenta and the constitutive relations take the form of an expansion in gradients of the charge densities.

The fluid variables of an uncharged relativistic fluid on a four-dimensional background $g_{(0)\m\n}$ are the components of its 4-momentum $\le<T^{t\m}(x)\ri>$ whose dynamics is governed by the covariant conservation of stress-energy,
\begin{align}
	\nabla_\m\le<T^{\m\n}(x)\ri>=0\;. \label{fluidEOM}
\end{align}
To zeroth order in gradients $\mathcal{O}(\pa^0)$, interactions between patches of local equilibrium are neglected and we are dealing with an ideal fluid
\begin{align}
	\le<T^{\m\n}(x)\ri>=\le[\e(x)+p(x)\ri]u^\m(x)u^\n(x)+p(x) g_{(0)}^{\m\n}(x) +\mathcal{O}\le(\pa\ri)\; \label{constituent0}
\end{align}
with local 4-velocity $u^\m$, energy density $\e$, and pressure $p$. The local equilibrium quantities $\e(x)$ and $p(x)$ are linked by the fluid's equation of state and satisfy the usual thermodynamic relations. We can invert the constitutive relation \eqref{constituent0} to use $\e$ and $u^\m$ as fluid variables instead of the 4-momentum.

The constitutive relations for an uncharged relativistic fluid are known up to third order in gradients \cite{Baier:2007ix,Romatschke:2009kr,Grozdanov:2015kqa}. Up to second order, seventeen independent tensor structures can be constructed from the fluid variables and $g_{(0)\m\n}$, each multiplied by a transport coefficient such as shear viscosity (see appendix \ref{AppendixConstituent} for the explicit expressions). These transport coefficients are the free input parameters of the effective hydrodynamic description. In order to compute their values we have to match the hydro result for appropriate correlators of $\le<T^{\m\n}\ri>$ with the corresponding result in the underlying microscopic theory.

\subsection{Sourced fluid stress tensor and Kubo formulae}\label{hydroResponse}

A suitable quantity to match is the response of $\le<T^{\m\n}\ri>$ to an external metric perturbation around flat space of a fluid in equilibrium. In the equilibrium rest frame, the fluid variables of the perturbed fluid on the background $g_{(0)\m\n}(x)=\eta_{\m\n}+h_{\m\n}(x)$ will take the form
\begin{align}
	\e(x)=\bar{\e}+\d\e(x)\;,&&u^\m(x)=\le(1,\underline{v}\ri)\le(-g_{(0)tt}-2g_{(0)ti}v^i-g_{(0)ij}v^iv^j\ri)^{-1/2}\;.
\end{align}
It is convenient to use $\d\e(x)$ and $\underline{v}(x)$ as fluid variables as this allows for a second expansion in fluctuations around static global equilibrium sourced by $h_{\r\s}$, in addition to the hydro gradient expansion. Writing the equilibrium stress tensor as
\begin{align}
	\bar{T}^{\m\n}\equiv\le<T^{\m\n}\ri>\le[\d\e=\underline{v}=h_{\r\s}=0\ri]\;,
\end{align}
the off-shell stress tensor of the perturbed fluid, eq.~\eqref{constituent2}, assumes the following form to first order $\mathcal{O}(\d)$ in fluctuations:
\begin{align}
	\le<T^{\m\n}\ri>\le[\d\e,\underline{v};h_{\r\s}\ri]=\bar{T}^{\m\n}+\le[\frac{\pa\bar{T}^{\m\n}}{\pa\,\d\e}\d\e+\frac{\pa\bar{T}^{\m\n}}{\pa v^i}v^i\ri]+\frac{\pa\bar{T}^{\m\n}}{\pa h_{\r\s}} h_{\r\s}+\mathcal{O}(\d^2)\;,
\end{align}
where we defined
\begin{align}
	\frac{\pa\bar{T}^{\m\n}}{\pa\,\d\e} \equiv \le.\frac{\pa\le<T^{\m\n}\ri>}{\pa\,\d\e}\ri |_{\d\e=\underline{v}=h_{\r\s}=0}
\end{align}
etc. Linearising the conservation eq.~\eqref{fluidEOM} around equilibrium yields the equations of motion for the fluid variables $\d\e$ and $\underline{v}$ in the presence of the linear metric perturbation $h_{\r\s}$. Defining
\begin{align}
	\d T^{\m\n}_{(\d\e,\underline{v})} \equiv \frac{\pa\bar{T}^{\m\n}}{\pa\,\d\e}\d\e+\frac{\pa\bar{T}^{\m\n}}{\pa v^i}v^i\;,&& \d T^{\m\n}_{(h)} \equiv \frac{\pa\bar{T}^{\m\n}}{\pa h_{\r\s}} h_{\r\s}\;,
\end{align}
these equations read
\begin{align}
	\pa_\m \d T^{\m\n}_{(\d\e,\underline{v})}=-\pa_\m \d T^{\m\n}_{(h)}-\d\Gamma^\m_{\m\r}\bar{T}^{\r\n}-\d\Gamma^\n_{\m\r}\bar{T}^{\m\r}+\mathcal{O}(\d^2) \label{linHydro}\;.
\end{align}
As $h_{\r\s}$ sources hydro fluctuations around static equilibrium we impose the boundary condition that $\d\e=\underline{v}=0$ for $h=0$, i.e.~we do not consider the usual free hydro modes that solve eq.~\eqref{linHydro} in the absence of the source terms on the right hand side. There exists a particularly simple subset of non-trivial metric perturbations in four dimensions $x^\m=(t,x,y,z)$ that do not source any fluctuations of the fluid variables to first order. Explicitly, if we only turn on
\begin{align}
	\le\{h_{xy}(t,z),h_{tx}(z),h_{ty}(z),h_{xz}(t),h_{yz}(t)\ri\} \label{bdyPert}
\end{align}
then the right hand side of eq.~\eqref{linHydro} vanishes at $\mathcal{O}(h)$, and the hydro equations are solved by $\d\e(h),\underline{v}(h)=\mathcal{O}(h^2)$. In that case, the \emph{on-shell} stress tensor takes the following form to second order in the source $\mathcal{O}(h^2)$:
\begin{align}
	\le<T^{\m\n}\ri>\le[h_{\r\s}\ri]&=\bar{T}^{\m\n}+\frac{\pa\bar{T}^{\m\n}}{\pa h_{\r\s}}h_{\r\s}+\frac{1}{2}\frac{\pa^2\bar{T}^{\m\n}}{\pa h_{\r\s}\pa h_{\kappa\l}}h_{\r\s}h_{\kappa\l}\nn\\
	&\quad+\le(\frac{\pa\bar{T}^{\m\n}}{\pa\,\d\e}\d\e(h)+\frac{\pa\bar{T}^{\m\n}}{\pa v^i}v^i(h)\ri)+\mathcal{O}(h^3)\;.
\end{align}
This expression simplifies even further if we focus on the transverse-tensor component $\le<T^{xy}(t,z)\ri>$ as it is independent of the scalars $\d\e$, $v^z$ and the transverse-vector components $v^x$, $v^y$ to first order $\mathcal{O}(\d)$:
\begin{align}
	\frac{\pa\bar{T}^{\m\n}}{\pa\,\d\e}=\frac{\pa\bar{T}^{\m\n}}{\pa v^i}=0\;.
\end{align}
Using this together with the constitutive relation \eqref{constituent2}, the on-shell response of $\le<T^{xy}\ri>$ to the metric perturbations \eqref{bdyPert} at $\mathcal{O}(h^2)$ is found to be
\begin{align}
	\le<T^{xy}\ri>&=\le[-\bar{p}-\eta\,\pa_t-\frac{\kappa}{2}\pa_z^2+\le(\eta\,\tau_\pi-\frac{\kappa}{2}+\kappa^*\ri)\pa_t^2\ri]h_{xy}(t,z)\nn\\
	&\quad+ \le[\bar{p}\,h_{xz}\,h_{yz}+\eta\le(h_{xz}\,\pa_th_{yz}+\pa_th_{xz}\,h_{yz}\ri)+\le(\l_1-\eta\,\tau_\pi-\frac{\kappa^*}{2}\ri)\pa_th_{xz}\,\pa_th_{yz} \ri.\nn\\
	&\quad\le.\quad +\le(\frac{\kappa}{2}-\eta\,\tau_\pi-\kappa^*\ri)\le(h_{xz}\,\pa_t^2h_{yz}+\pa_t^2h_{xz}\,h_{yz}\ri)\ri] \nn\\
	&\quad+\le[-\bar{p}\,h_{tx}\,h_{ty}+\le(\frac{\l_3}{4}-\frac{\kappa^*}{2}\ri)\,\pa_zh_{tx}\,\pa_zh_{ty}-\frac{\kappa}{2}\le(h_{tx}\,\pa_z^2h_{ty}+\pa_z^2h_{tx}\,h_{ty}\ri) \ri] \nn\\
	&\quad+\le[\frac{1}{2}\eta\,\tau_\pi-\frac{\l_2}{4}+\frac{\kappa^*}{2}\ri]\le(\pa_zh_{tx}\,\pa_th_{yz}+\pa_zh_{ty}\,\pa_th_{xz}\ri) +\mathcal{O}(h^3,\pa^3)\;, \label{Txyresponse}
\end{align}
where $\bar{p}$ denotes the pressure in global equilibrium. The linear response sourced by the tensor perturbation $h_{xy}$ was derived in ref.~\cite{Baier:2007ix}, the quadratic response sourced by the transverse-vector perturbations was computed in ref.~\cite{Moore:2010bu} for a conformal fluid with $\kappa^*=0$. To our knowledge, the response \eqref{Txyresponse} of a \emph{non-conformal} fluid has not appeared in the literature before. Note in particular that $\kappa^*$ is the only non-conformal second-order coefficient that appears in eq.~\eqref{Txyresponse}. Eq.~\eqref{Txyresponse} shows that the response of $\le<T^{xy}\ri>$ to the perturbations \eqref{bdyPert} gives us access to five independent linear combinations of second-order transport coefficients\footnote{Note that if we wanted to extract all fifteen second-order coefficients we would have to turn on metric perturbations in the scalar sound channel which would necessarily source fluctuations of $\d\e$ and $\underline{v}$.},
\begin{align}
	\kappa\;,\;\;\;\,\eta\,\tau_\pi+\kappa^*\;,\;\;\;\l_1+\frac{\kappa^*}{2}\;,\;\;\;\l_2\;,\;\;\;\l_3-2\kappa^*\;,
\end{align}
which includes the combination $H=2\eta\,\tau_\pi-4\l_1-\l_2$. In fact, no additional information is provided by the response to $h_{xy}$, and all five coefficients \eqref{coeffs} can be obtained by turning on plane-wave excitations for $\le\{h_{xz}(t),h_{yz}(t)\ri\}$, $\le\{h_{tx}(z),h_{ty}(z)\ri\}$, and $\le\{h_{ty}(z),h_{xz}(t)\ri\}$, one after another.

\paragraph{Turning on $\le\{h_{xz}(t),h_{yz}(t)\ri\}$:}

Perturbing the metric by
\begin{align}
	\frac{1}{2}h_{\m\n}\dif x^\m\dif x^\n=\e\le(H_{xz}^{(b)}e^{-iq_0t}\dif x\dif z+H_{yz}^{(b)}e^{-ip_0t}\dif y\dif z\ri)\;, \label{h1}
\end{align}
with plane-wave amplitudes $H_{xz}^{(b)}$ and $H_{yz}^{(b)}$, sources the response
\begin{align}
	\le<T^{xy}(x)\ri>&=\le[\bar{p}-i\le(q_0+p_0\ri)\eta-q_0p_0\le(\l_1-\eta\,\tau_\pi-\frac{\kappa^*}{2}\ri) \ri.\nn\\
	&\quad\le.-\le(q_0^2+p_0^2\ri)\le(\frac{\kappa}{2}-\eta\,\tau_\pi-\kappa^*\ri)\ri]\e^2H_{xz}^{(b)}H_{yz}^{(b)}e^{-i\le(q_0+p_0\ri)t}+\mathcal{O}(\e^3,\pa^3)\;,\label{Txy1}
\end{align}
corresponding to the second and third line in eq.~\eqref{Txyresponse}.

\paragraph{Turning on $\le\{h_{tx}(z),h_{ty}(z)\ri\}$:}

Perturbing the metric by
\begin{align}
	\frac{1}{2}h_{\m\n}\dif x^\m\dif x^\n=\e\le(H_{tx}^{(b)}e^{iq_zz}\dif t\dif x+H_{ty}^{(b)}e^{ip_zz}\dif t\dif y\ri)\;, \label{h2}
\end{align}
with plane-wave amplitudes $H_{tx}^{(b)}$ and $H_{ty}^{(b)}$, sources the response
\begin{align}
	\le<T^{xy}(x)\ri>&=\le[-\bar{p}-q_zp_z\le(\frac{\l_3}{4}-\frac{\kappa^*}{2}\ri)+\le(q_z^2+p_z^2\ri)\frac{\kappa}{2} \ri] \e^2H_{xz}^{(b)}H_{yz}^{(b)}e^{i\le(q_z+p_z\ri)z}+\mathcal{O}(\e^3,\pa^3)\;, \label{Txy2}
\end{align}
corresponding to the fourth line in eq.~\eqref{Txyresponse}.

\paragraph{Turning on $\le\{h_{ty}(z),h_{xz}(t)\ri\}$:}

Perturbing the metric by
\begin{align}
	\frac{1}{2}h_{\m\n}\dif x^\m\dif x^\n=\e\le(H_{ty}^{(b)}e^{ip_zz}\dif t\dif y+H_{xz}^{(b)}e^{-iq_0t}\dif x\dif z\ri)\;, \label{h3}
\end{align}
with plane-wave amplitudes $H_{ty}^{(b)}$ and $H_{xz}^{(b)}$, sources the response
\begin{align}
	\le<T^{xy}(x)\ri>&=q_0p_z\le(\frac{1}{2}\eta\,\tau_\pi-\frac{\l_2}{4}+\frac{\kappa^*}{2}\ri)\e^2H_{ty}^{(b)}H_{xz}^{(b)}e^{-iq_0t+ip_zz}+\mathcal{O}(\e^3,\pa^3)\;, \label{Txy3}
\end{align}
corresponding to the last line in eq.~\eqref{Txyresponse}.

\paragraph{Kubo formulae}

Comparing eqs.~\eqref{Txy1}, \eqref{Txy2}, \eqref{Txy3} with the general form of the stress-tensor response written in terms of retarded correlators in momentum space \cite{Moore:2010bu},
\begin{align}
	\le<T^{\m\n}(x=0)\ri>&=G^{\m\n}(0)-\frac{1}{2}\int\frac{\dif^4p}{\le(2\pi\ri)^4}G^{\m\n,\r\s}(p)h_{\r\s}(p)\nn\\
	&\quad+\frac{1}{8}\int\frac{\dif^4q}{\le(2\pi\ri)^4}\frac{\dif^4p}{\le(2\pi\ri)^4}G^{\m\n,\r\s,\kappa\l}(q,p)h_{\r\s}(q)h_{\kappa\l}(p)+\mathcal{O}(h^3)\;,
\end{align}
one can read off the low-momentum expansion of the corresponding three-point functions $G^{xy,xz,yz}$, $G^{xy,tx,ty}$, $G^{xy,ty,xz}$ and derive the following Kubo formulae\footnote{Ref.~\cite{Moore:2010bu} defines the Fourier-transformed three-point functions with the opposite sign for the two momenta. In our convention, the shear viscosity is therefore given by $\eta=i\le.\pa_{q_0}G^{xy,xz,yz}(q,p)\ri|_{q=p=0}$, as opposed to eq.~(21) in ref.~\cite{Moore:2010bu}, $\eta=-i\le.\pa_{q_0}G^{xy,xz,yz}(q,p)\ri|_{q=p=0}$. We further believe that the factors of 2 in their eqs.~(22) and (23) should be absent, in agreement with their eq.~(26).}:
\begin{subequations}\label{Kubo}
\begin{align}
	\kappa&=\le.\pa_{q_z}^2G^{xy,tx,ty}(q,p)\ri|_{q=p=0}\;,\\
	\eta\tau_\pi+\kappa^*&=\frac{\kappa}{2}+\frac{1}{2}\le.\pa_{q_0}^2G^{xy,xz,yz}(q,p)\ri|_{q=p=0}\;,\\
	\l_1+\frac{\kappa^*}{2}&=\le(\eta\tau_\pi+\kappa^*\ri)-\le.\pa_{q_0}\pa_{p_0}G^{xy,xz,yz}(q,p)\ri|_{q=p=0}\;,\\
	\l_2&=2\le(\eta\tau_\pi+\kappa^*\ri)-4\le.\pa_{q_0}\pa_{p_z}G^{xy,tx,xz}(q,p)\ri|_{q=p=0}\;,\\
	\l_3-2\kappa^*&=-4\le.\pa_{q_z}\pa_{p_z}G^{xy,tx,ty}(q,p)\ri|_{q=p=0}\;.
\end{align}
\end{subequations}

\subsection{Holographic calculation}\label{HologComput}

We now turn to the specific kind of microscopic theories that we will be studying throughout the remainder of this work: strongly coupled non-conformal QFTs with a holographic dual in asymptotically $AdS_5$ \cite{Aharony:1999ti,Ammon:2015wua,Zaanen:2015oix}. In order to extract their second-order transport coefficients we need to compute the stress-tensor component $\le<T^{xy}\ri>$ to second order $\mathcal{O}(\e^2)$ in the perturbations \eqref{h1}, \eqref{h2}, \eqref{h3} of the field-theory metric and match the result with the effective hydro results \eqref{Txy1}, \eqref{Txy2}, \eqref{Txy3} \cite{Son:2007vk,Saremi:2011nh,Arnold:2011ja}. Perturbations of the external field-theory metric act as boundary sources for perturbations of the dynamical bulk metric $g_{mn}$ in the dual gravity theory. Their backreaction on the bulk can be computed perturbatively in $\e$. With regard to universality it is encouraging that, even for non-conformal fluids, $H$ can be measured by considering shear perturbations of the fluid. Unlike sound perturbations, shear perturbations only couple to the gravity sector of the dual bulk, which is common to all holographic theories, and not to the model-specific matter content \cite{Benincasa:2005iv}. 
 
According to the holographic dictionary \cite{Gubser:1998bc,Witten:1998qj}, the field-theory stress tensor $\le<T^{xy}\ri>$ equals, up to a scaling factor, the quasi-local gravity stress tensor $\mathcal{T}^{\m\n}$ of the dual AdS bulk \cite{Balasubramanian:1999re} (see appendix \ref{renormalisation}). The latter measures the response of the on-shell gravity action to changes in the induced AdS boundary metric. Explicitly, variations of the induced AdS boundary metric $\g_{\m\n}$ lead to the following variation in the (appropriately renormalised) bulk action,
\begin{align}
	\d S^\mathrm{ren}=-\frac{1}{16\pi G_N}\int\dif^5x\sqrt{-g}\,\mathrm{EOM}^{mn}\d g_{mn}+\frac{1}{2}\int\limits_{\pa AdS_5}\dif^4x\sqrt{-\g}\,\mathcal{T}^{\m\n}\d\g_{\m\n}\;,
\end{align}
where $\mathrm{EOM}_{mn}$ denote Einstein's equations in the bulk. Thus, for $\le(2/\sqrt{-\g}\ri)\le(\d S^\mathrm{ren}/\d\g_{xy}\ri)$ to yield the correct result for $\mathcal{T}^{xy}$ up to $\mathcal{O}(\e^2)$ included, $\mathrm{EOM}^{xy}$ must be satisfied to order $\mathcal{O}(\e^2)$ \cite{Saremi:2011nh}.\footnote{Given that the bulk metric will be diagonal to leading order, $g_{mn}\propto \d_{mn}$, we can ensure that $\mathrm{EOM}^{xy}=\mathcal{O}(\e^3)$ by solving the usual form of Einstein's equations with lower indices, $\mathrm{EOM}_{mn}$, to $\mathcal{O}(\e)$ and $\mathrm{EOM}_{xy}$ to $\mathcal{O}(\e^2)$.}

To summarise, our strategy to compute the transport coefficients will be the following: we turn on the boundary metric perturbations \eqref{h1}, \eqref{h2}, \eqref{h3}, one after another, solve the $xy$-component of Einstein's equations in the bulk to second order in amplitudes $\mathcal{O}(\e^2)$ and momenta $\mathcal{O}(\pa^2)$ of the perturbations, compute the $xy$-component of the field-theory stress tensor according to the holographic dictionary, and finally compare it with the general hydro results \eqref{Txy1}, \eqref{Txy2}, \eqref{Txy3}.


\section{A class of non-conformal holographic models}\label{model}

In this section we introduce the specific class of strongly coupled non-conformal field theories, described by a holographic gravity dual, whose transport properties we are going to study: we will consider holographic RG flows \cite{Girardello:1998pd,Girardello:1999hj,Girardello:1999bd,Freedman:1999gp,Pilch:2000ue,Klebanov:2000hb,Maldacena:2000yy,Bianchi:2001de,Bigazzi:2003ui,Buchel:2003ah} which, starting from a four-dimensional conformal field theory (CFT) in the UV, are triggered by a relevant deformation
\begin{align}
	\int\dif^4x\sqrt{-g_{(0)}}\,\Lambda(x)O(x)
\end{align}
with a scalar operator $O$ of dimension $\D=3$.\footnote{We restrict ourselves to operators of dimension $\D=3$ because the holographic renormalisation has already been done for this class of holographic RG flows, see appendix \ref{renormalisation} for details.} The dual gravity description of such RG flows is provided by five-dimensional Einstein gravity coupled to a scalar field,
\begin{align}
	S=\frac{1}{16\pi G_N}\int\dif^5x\sqrt{-g}\le[R-\frac{1}{2}\le(\pa\f\ri)^2-V(\f)\ri]\;, \label{action}
\end{align}
with potentials of the form
\begin{align}
	V(\f)=\frac{1}{L^2}\le[-12-\frac{3}{2}\f^2+\mathcal{O}(\f^4)\ri]\;.\label{potential}
\end{align}
Solutions to the bulk equations of motion approach $AdS_5$ with radius $L$ in the near-boundary region $\f\rightarrow0$,
\begin{align}
	\dif s^2\rightarrow\frac{L^2}{\z^2}\le(\dif \z^2+\dif x\cdot\dif x\ri)\;,\label{zGauge}
\end{align}
which is dual to the UV fixed point. The leading near-boundary mode of the scalar with mass $m^2L^2=\D\le(\D-4\ri)=-3$ is $\f\sim\Lambda \z$, where $\Lambda$ is interpreted as the source of the dual operator $O$.

\subsection{Background equations of motion}\label{backgroundEqs}

Thermal equilibrium states in flat space are holographically described by black-brane solutions to \eqref{action} that preserve Euclidean symmetry in the spatial directions \cite{Maldacena:1997re,Witten:1998zw}. Choosing a convenient gauge for the radial coordinate, they can be written as
\begin{align}
	\dif s^2=g_{mn}^{(0)}\dif x^m\dif x^n=e^{2A(u)}\le[-f(u)\dif t^2+\dif\underline{x}^2\ri]+\frac{L^2}{4u^2f(u)}\dif u^2\;,\label{uGauge}
\end{align}
where $f(u)$ has a simple zero at the horizon and the Hawking temperature $T$ and entropy density $s$ are given by
\begin{align}
	T=\le.\frac{-f'(u)\,e^{A(u)}}{2\pi L}\ri|_{u=1}\;,&&s=\le.\frac{e^{3A(u)}}{4G_N}\ri|_{u=1}\;.\label{Hawking}
\end{align}
The residual scaling symmetry, inherited from the UV CFT, can be used to set the horizon position to $u=1$, effectively expressing all dimensionful quantities in units of temperature. The equations of motion that follow from the action \eqref{action} take the following form in the gauge \eqref{uGauge}:
\begin{subequations}
\label{uEOM}
\begin{align}
	\f''+\le(4A'+\frac{1}{u}\ri)\f'+\frac{f'}{f}\f'-\frac{L^2}{4u^2f}\le(\frac{\dif V}{\dif \phi}\ri)&=0\;,\label{uEOM1}\\
	A''+\frac{1}{u}A' +\frac{1}{6}\le(\f'\ri)^2&=0\;,\label{uEOM2}\\
	f''+\le(4A'+\frac{1}{u}\ri)f'&=0\;,\label{uEOM3}\\
	6A' f' +f\le[24\le(A'\ri)^2- \le(\f'\ri)^2\ri]+\frac{L^2}{2u^2}V&=0\;,\label{uConstraint}
\end{align}
\end{subequations}
where primes denote derivatives with respect to $u$. The system is partly redundant in the sense that the constraint \eqref{uConstraint} and its derivative are algebraically given in terms of the other three equations:
\begin{align}
	\le(\frac{\dif}{\dif u}+\frac{2}{u}\ri)\eqref{uConstraint}=-2f\frac{\dif \f}{\dif u}\eqref{uEOM1}+\le(48 fA'+6f'\ri)\eqref{uEOM2}+6A'\eqref{uEOM3}  \;.\label{uRedundancy}
\end{align}
For vanishing scalar, $\f=0$, eqs.~\eqref{uEOM} are solved by the $AdS_5$-black brane background:
\begin{align}
	A(u)=\frac{1}{2}\log\le[\frac{\le(\pi TL\ri)^2}{u}\ri]\;,&&f(u)=1-u^2\;.\label{blackBranes}
\end{align}

\subsection{Equations for metric perturbations}\label{perturbEqs}

We will now present the dual gravity description of the (field-theory) metric perturbations we discussed in section \ref{hydroSection}. Generally, the external metric $g_{(0)\m\n}$ of QFTs with a holographic dual prescribes the value of the dual dynamical bulk metric $g_{mn}$ at the AdS boundary \cite{Witten:1998qj}. Perturbations $h_{\m\n}$ of the field-theory metric source changes in the bulk metric, which in turn encode the response of the QFT stress tensor $\le<T^{\m\n}\ri>$ \cite{Balasubramanian:1999re,deHaro:2000vlm}. Denoting the field-theory directions by $x^\m=(t,x,y,z)$ and maintaining a radial gauge $g_{u\m}=0$, we will write the perturbed bulk metric as
\begin{align}
	\dif s^2=g_{mn}\dif x^\m \dif x^n=g_{mn}^{(0)}\dif x^m\dif x^n+\e\,g_{\m\n}^{(1)}\dif x^\m\dif x^\n+\e^2\,g_{\m\n}^{(2)}\dif x^\m\dif x^\n+\mathcal{O}(\e^3) \;, \label{fullMetric}
\end{align}
where $g_{mn}^{(0)}$ is the background metric \eqref{uGauge}, $\e\,g_{\m\n}^{(1)}$ contains the sourced metric perturbations at the boundary, and $\e^2\,g_{\m\n}^{(2)}$ describes their $\mathcal{O}(\e^2)$ backreaction on the bulk. The form of $g_{\m\n}^{(1)}$ and $g_{\m\n}^{(2)}$ varies depending on which of the three metric perturbations \eqref{h1}, \eqref{h2} and \eqref{h3} we turn on. However, none of these transverse-vector perturbations source linear fluctuations of the scalar field at $\mathcal{O}(\e)$, and fluctuations of $\f$ at $\mathcal{O}(\e^2)$ do not affect $g_{xy}^{(2)}$.\footnote{Metric perturbations in the scalar sound channel, on the other hand, would source linear fluctuations of $\f$. It is for this technical reason that we restrict ourselves to perturbations in the transverse shear channel. The drawback is that this only gives us access to five independent combinations of transport coefficients, eq.~\eqref{coeffs}, as explained in section \ref{hydroResponse}.}

\paragraph{Turning on $\le\{h_{xz}(t),h_{yz}(t)\ri\}$:}

The metric perturbation \eqref{h1} corresponds to
\begin{align}
	\frac{1}{2}g_{\m\n}^{(1)}\dif x^\m\dif x^\n=e^{2A(u)}&\le[ H_{xz}^{(b)}e^{-iq_0t}\,H^{(1t)}(u,q_0)\,\dif x\dif z\ri.\nn\\
	&\le.+H_{yz}^{(b)}e^{-ip_0t}\,H^{(1t)}(u,p_0)\,\dif y\dif z\ri]\;,\label{case1Bulk}
\end{align}
where Einstein's equations at order $\mathcal{O}(\e)$ reduce to a frequency-dependent equation for the function $H^{(1t)}$,
\begin{align}
	H^{(1t)\prime\prime}(u,\omega)+\left(\frac{1}{u}+4A'+\frac{f'}{f}\right)H^{(1t)\prime}(u,\omega)+\frac{e^{-2A}L^2\,\omega^2}{4u^2f^2}H^{(1t)}(u,\omega)=0 \;, \label{H1tEOM}
\end{align}
normalised to $H^{(1t)}(u=0)=1$ at the boundary. The $xy$-component of the resulting backreaction at $\mathcal{O}(\e^2)$ is conveniently parameterised as
\begin{align}
	g_{xy}^{(2)}=e^{2A(u)}\,H_{xz}^{(b)}\,H_{yz}^{(b)}\,e^{-ip_0t-iq_0t}H^{(2tt)}(u,q_0,p_0)
\end{align}
so that the $xy$-component of Einstein's equations at $\mathcal{O}(\e^2)$ becomes
\begin{align}
	&H^{(2tt)\prime\prime}(u,q_0,p_0)+\left(\frac{1}{u}+4A'+\frac{f'}{f}\right)H^{(2tt)\prime}(u,q_0,p_0)+\frac{e^{-2A}L^2(q_0+p_0)^2}{4u^2f^2}H^{(2tt)}(u,q_0,p_0)\nn\\
	&=\frac{e^{-2A}L^2 q_0 p_0}{4u^2 f^2}H^{(1t)}(u,q_0)H^{(1t)}(u,p_0)+H^{(1t)\prime}(u,q_0)H^{(1t)\prime}(u,p_0)\;.
	\label{H2ttEOM}
\end{align}
The fluctuation $H^{(2tt)}$ is not sourced by an explicit deformation of the boundary metric, $H^{(2tt)}(u=0)=0$, but only by the backreaction of the first-order perturbations $g_{\m\n}^{(1)}$.

\paragraph{Turning on $\le\{h_{tx}(z),h_{ty}(z)\ri\}$:}

The metric perturbation \eqref{h2} corresponds to
\begin{align}
	\frac{1}{2}g_{\m\n}^{(1)}\dif x^\m\dif x^\n=e^{2A(u)}\le[H_{tx}^{(b)}e^{iq_zz}\,H^{(1z)}(u,q_z)\,\dif t\dif x+H_{ty}^{(b)}e^{ip_zz}\,H^{(1z)}(u,p_z)\,\dif t\dif y\ri]\;,\label{case2Bulk}
\end{align}
where Einstein's equations at order $\mathcal{O}(\e)$ reduce to a momentum-dependent equation for the function $H^{(1z)}$,
\begin{align}
	H^{(1z)\prime\prime}(u,\omega)+\left(\frac{1}{u}+4A'\right)H^{(1z)\prime}(u,\omega)-\frac{e^{-2A}L^2\,\omega^2}{4u^2f}H^{(1z)}(u,\omega)=0 \;, \label{H1zEOM}
\end{align}
normalised to $H^{(1z)}(u=0)=1$ at the boundary. The $xy$-component of the resulting backreaction at $\mathcal{O}(\e^2)$ is conveniently parameterised as
\begin{align}
	g_{xy}^{(2)}=e^{2A(u)}\,H_{tx}^{(b)}\,H_{ty}^{(b)}\,e^{iq_zz+ip_zz}H^{(2zz)}(u,q_z,p_z)
\end{align}
so that the $xy$-component of Einstein's equations at $\mathcal{O}(\e^2)$ becomes
\begin{align}
	&H^{(2zz)\prime\prime}(u,q_z,p_z)+\left(\frac{1}{u}+4A'+\frac{f'}{f}\right)H^{(2zz)\prime}(u,q_z,p_z)-\frac{e^{-2A}L^2\,\le(q_z+p_z\ri)^2}{4u^2f}H^{(2zz)}(u,q_z,p_z) \nn\\
	&=\frac{e^{-2A}L^2 q_z p_z}{4u^2 f^2}H^{(1z)}(u,q_z)H^{(1z)}(u,p_z)-\frac{1}{f}H^{(1t)\prime}(u,q_z)H^{(1z)\prime}(u,p_z)\;.
	\label{H2zzEOM}
\end{align}
The fluctuation $H^{(2zz)}$ is not sourced by an explicit deformation of the boundary metric, $H^{(2zz)}(u=0)=0$.

\paragraph{Turning on $\le\{h_{ty}(z),h_{xz}(t)\ri\}$:}

The metric perturbation \eqref{h3} corresponds to
\begin{align}
	\frac{1}{2}g_{\m\n}^{(1)}\dif x^\m\dif x^\n=e^{2A(u)}\le[H_{ty}^{(b)}e^{ip_zz}\,H^{(1z)}(u,p_z)\,\dif t\dif y+H_{xz}^{(b)}e^{-iq_0t}\,H^{(1t)}(u,q_0)\,\dif x\dif z\ri]\;,\label{case3Bulk}
\end{align}
where Einstein's equations at order $\mathcal{O}(\e)$ reduce to eqs.~\eqref{H1tEOM} and \eqref{H1zEOM} for $H^{(1t)}$ and $H^{(1z)}$ respectively.\footnote{Moreover, $H^{(1t)}$ and $H^{(1z)}$ are again normalised to 1 at the boundary and, as we will discuss in section \ref{localPerturb}, are subject to the same boundary conditions at the horizon as they are in cases \eqref{case1Bulk} and \eqref{case2Bulk}. Hence $H^{(1t)}$ and $H^{(1z)}$ here are indeed the same as in perturbations \eqref{case1Bulk} and \eqref{case2Bulk}.} The $xy$-component of the resulting backreaction at $\mathcal{O}(\e^2)$ is conveniently parameterised as
\begin{align}
	g_{xy}^{(2)}=e^{2A(u)}\,H_{ty}^{(b)}\,H_{xz}^{(b)}\,e^{-iq_0t+ip_zz}H^{(2tz)}(u,q_0,p_z)
\end{align}
so that the $xy$-component of Einstein's equations at $\mathcal{O}(\e^2)$ becomes
\begin{align}
	&H^{(2tz)\prime\prime}(u,q_0,p_z)+\left(\frac{1}{u}+4A'+\frac{f'}{f}\right)H^{(2tz)\prime}(u,q_0,p_z) \nn\\
	&-\frac{e^{-2A}L^2\,\le(-q_0^2+f\,p_z^2\ri)}{4u^2f^2}H^{(2tz)}(u,q_0,p_z) =\frac{e^{-2A}L^2 q_0 p_z}{4u^2 f^2}H^{(1t)}(u,q_0)H^{(1z)}(u,p_z)\;.
	\label{H2tzEOM}
\end{align}
The fluctuation $H^{(2tz)}$ is not sourced by an explicit deformation of the boundary metric, $H^{(2tz)}(u=0)=0$.


\section{Solving Einstein's equations}\label{einstein}

In the previous section we introduced a class of non-conformal holographic models and presented the corresponding equations of motion for static black-brane backgrounds (subsection \ref{backgroundEqs}) and for metric fluctuations around these backgrounds (subsection \ref{perturbEqs}), sourced by perturbations \eqref{h1}, \eqref{h2}, \eqref{h3} of the field-theory metric. Our next goal is to determine the response of the field-theory stress tensor $\le<T^{\m\n}\ri>$ to these perturbations. To this end, we need to find solutions to the bulk equations of motion that satisfy the appropriate boundary conditions at the AdS boundary and at the horizon. The stress tensor is then encoded in the near-boundary expansion of the bulk metric \cite{Myers:1999psa,deHaro:2000vlm}.

In this section we want to see how far we can get solving for fluctuations of the bulk metric around an \emph{arbitrary} background solution without specifying the scalar potential \eqref{potential} beyond the mass term. The results of this section therefore apply to all holographic RG flows triggered by a scalar operators of dimension  $\D=3$, at any value of the temperature. We begin by writing down the near-horizon and near-boundary expansion of a general background solution in subsection \ref{backgroundEinstein}. The former is needed to identify which boundary conditions to impose on metric fluctuations at the horizon while the latter is necessary for the computation of $\le<T^{\m\n}\ri>$. In subsection \ref{localPerturb} we turn to fluctuations of the bulk metric: we impose the appropriate boundary conditions, perform the hydro expansion up to second order in momenta, determine local solutions near horizon and boundary, and try to find global solutions connecting the two. We managed to solve for all but five of the hydro metric fluctuations analytically, and we found integral expressions for another four.\footnote{Curiously, it turns out that the one hydro metric fluctuation for which we did not find a solution cancels out in the expression for $\le<T^{\m\n}\ri>$ presented in subsection \ref{FormulaeCoefficients}.}

We appreciate that this is a rather technical section. For practical purposes, integrals \eqref{Yintegral}, which yield the sub-leading boundary modes for four hydro metric fluctuations, make up the central result we will refer to in subsequent sections.


\subsection{Local analysis of background solutions}\label{backgroundEinstein}
\paragraph{Near-horizon expansion}
The fields $A(u)$, $f(u)$ and $\phi(u)$ satisfy a system of three second-order equations \eqref{uEOM1}--\eqref{uEOM3} and one first-order equation \eqref{uConstraint}. The coefficients in the local series solution around the horizon are not constrained by \eqref{uConstraint} thanks to the redundancy \eqref{uRedundancy}. The general local near-horizon solution therefore contains six integration constants. Demanding that $A$ and $\phi$ be regular at the horizon and that the horizon position be at $u=1$ amounts to three boundary conditions. The near-horizon expansions of $A$, $f$, and $\phi$ thus depend on three near-horizon modes $\le\{A_H,f_H,\f_H\ri\}$ and assume the form
\begin{subequations}
\label{NHu}
\begin{align}
	A(u)&=A_H+\sum\limits_{k\geq1}b_k^{A}\le(1-u\ri)^k\;,\\
	f(u)&=\le(1-u\ri)\le[f_H+\sum\limits_{k\geq1}b_k^{f}\le(1-u\ri)^k\ri]\;,\\
	\f(u)&=\f_H+\frac{L^2\,V'(\f_H)}{4f_H}\le(1-u\ri)+\sum\limits_{k\geq2}b_k^{\f}\le(1-u\ri)^k\;,\label{phiNH}
\end{align}
\end{subequations}
with series coefficients fully determined by the near-horizon modes and the given potential.
\paragraph{Near-boundary expansion}
The general local near-boundary solution to \eqref{uEOM} contains six integration constants, two of which we fix by demanding that the spacetime be asymptotically $AdS_5$,
\begin{align}
	A(u)&=-\frac{1}{2}\log(u)+\mathcal{O}\le(u^0\ri)\;,&&f(u)=1+\mathcal{O}(u)\;.
\end{align}
The redundancy \eqref{uRedundancy} implies that at each order in $u$, the four equations of motion \eqref{uEOM} only provide three independent algebraic equations which determine the corresponding series coefficients of $A$, $f$ and $\f$. When we reach the order of one of the remaining four integration constants (the sub-leading modes of $A$ and $f$, and the two modes of the scalar $\f$), then the three algebraic equations fail to fix these (free) parameters. Instead, either only two of the algebraic equations are independent at this order, or one of them restricts the form of potentials $V$, eq.~\eqref{potential}, compatible with the requirement of asymptotically $AdS_5$.

One finds that near-boundary solutions take the form
\begin{subequations}
\label{NBu}
\begin{align}
	A(u)&=\frac{1}{2}\log\le(\frac{A_b}{u}\ri)-\frac{\f_L^2}{24}u+\sum\limits_{k\geq2}c_k^{A}u^k\;,\label{ANB}\\
	f(u)&=1+f_b\,u^2+u^2\,\sum\limits_{k\geq1}c_k^{f}u^k\;,\\
	\f(u)&=\f_L\,\sqrt{u}+\f_{SL}\,u^{3/2}+u^{3/2}\,\sum\limits_{k\geq1}c_k^{\f}u^k\;,
\end{align}
\end{subequations}
with series coefficients fully determined by $V$ and the four boundary modes $\le\{A_b,f_b,\f_L,\f_{SL}\ri\}$, and that the potential must satisfy\footnote{Note that superpotentials $W$ which lead to $L^2\,V=-12-\le(3/2\ri)\f^2+\mathcal{O}\le(\f^4\ri)$, i.e.~$L\,W=-\le(3/2\ri)-\f^2/8+\mathcal{O}\le(\f^4\ri)$, automatically ensure that the condition \eqref{Vcondition} is satisfied. We believe that this condition was overlooked in ref.~\cite{Gubser:2008ny}.}
\begin{align}
	4!\,\frac{\dif^4V}{\dif\f^4}=-\frac{1}{12L^2}\;.\label{Vcondition}
\end{align}
As the local near-boundary solution cannot depend on more than four free parameters, the three independent equations coming from \eqref{uEOM} must succeed in fixing the three corresponding series coefficients of $A$, $f$, and $\f$ to all orders beyond the fields' sub-leading modes. In particular, no further constraints on the potential $V(\f)$ can arise.

\subsection{Solutions for metric perturbations}\label{localPerturb}

\paragraph{Boundary conditions}

In order to compute the retarded response of the stress tensor, time-dependent perturbations of the bulk metric need to represent incoming waves at the horizon \cite{Son:2002sd,Herzog:2002pc} while static perturbations are simply required to be regular at the horizon.  Let us define momenta in Fraktur as the dimensionless combination\
\begin{align}
	\mathfrak{w}\equiv\frac{L\,\omega}{2f_He^{A_H}}=\frac{\omega}{4\pi T}\;, \label{fraktur}
\end{align}
where we used expression \eqref{Hawking} for the temperature. The incoming-wave solution to eq.~\eqref{H1tEOM} and the regular solution to eq.~\eqref{H1zEOM} take the form
\begin{subequations}\label{H1t1z}
\begin{align}
	H^{(1t)}(u,\omega)&=\le(1-u\ri)^{-i\mathfrak{w}}K^{(1t)}(u,\omega)\;,\\
	H^{(1z)}(u,\omega)&=\le(1-u\ri)K^{(1z)}(u,\omega)\;,
\end{align}
\end{subequations}
where $K^{(1\a)}$, $\a\in\le\{t,z\ri\}$, are analytic at the horizon and normalised to 1 at the boundary. The first-order perturbations dictate the form of the second-order fluctuations they source. Investigating eqs.~\eqref{H2ttEOM}, \eqref{H2zzEOM}, \eqref{H2tzEOM} shows that
\begin{subequations}\label{H2ttzztz}
\begin{align}
	H^{(2tt)}(u,q_0,p_0)&=\le(1-u\ri)^{-i\mathfrak{q}_0-i\mathfrak{p}_0}K^{(2tt)}(u,q_0,p_0)\;,\\
	H^{(2zz)}(u,q_z,p_z)&=K^{(2zz)}(u,q_z,p_z)\;,\\
	H^{(2tz)}(u,q_0,p_z)&=\le(1-u\ri)^{-i\mathfrak{q}_0}K^{(2tz)}(u,q_0,p_z)\;,
\end{align}
\end{subequations}
where $K^{(2\b)}$, $\b\in\le\{tt,zz,tz\ri\}$, are analytic at the horizon and vanish at the boundary.

\paragraph{Hydrodynamic gradient expansion}

To match the holographic result for $\le<T^{xy}\ri>$ with the general hydro form discussed in section \ref{hydroResponse}, we need to turn on sources that admit an expansion in small gradients/momenta. We therefore expand the metric fluctuations as
\begin{align}
	K^{(1\a)}(u,\omega)&=K_0^{(1\a)}(u)+K_1^{(1\a)}(u)\,\omega+K_2^{(1\a)}(u)\,\omega^2+\mathcal{O}(\mathfrak{w}^3)\;,\;\;\a\in\le\{t,z\ri\}\\
	K^{(2\b)}(u,q,p)&=K_{(0,0)}^{(2\b)}(u)+\le[K_{(1,0)}^{(2\b)}(u)\,q+K_{(0,1)}^{(2\b)}(u)\,p\ri]\nn\\
	&+\le[K_{(2,0)}^{(2\b)}(u)\,q^2+K_{(1,1)}^{(2\b)}(u)\,q\,p+K_{(0,2)}^{(2\b)}(u)\,p^2\ri]+\le(\mathcal{O}(\mathfrak{q},\mathfrak{p})\ri)^3\;,\;\;\b\in\le\{tt,zz,tz\ri\}\;.\nn
\end{align}
To simplify the discussion we shall use $a$ to label the metric perturbations and $j$ to denote the order in the hydro expansion, i.e.~$j\in\le\{0,1\dots\ri\}$ for $a\in\le\{1t,1z\ri\}$ and $j\in\le\{(0,0),(1,0),(0,1)\dots\ri\}$ for $a\in\le\{2tt,2zz,2tz\ri\}$. 

Expanding the equations of motion \eqref{H1tEOM}, \eqref{H1zEOM}, \eqref{H2ttEOM}, \eqref{H2zzEOM}, \eqref{H2tzEOM} in momenta reveals that not all of the $K^{(a)}_j$ are independent. Firstly, $K^{(2tt)}_{(1,0)}$ and $K^{(2tt)}_{(0,1)}$ satisfy the same equation and boundary conditions and are thus identical. As a consequence the same holds for $K^{(2tt)}_{(2,0)}$ and $K^{(2tt)}_{(0,2)}$. Furthermore, $K^{(1z)}_1$ satisfies a linear and homogeneous equation. The solution that is regular at the horizon and vanishes at the boundary is identically zero. The same is then true for $K^{(2zz)}_{(1,0)}$, $K^{(2zz)}_{(0,1)}$, $K^{(2tz)}_{(0,0)}$, $K^{(2tz)}_{(1,0)}$, $K^{(2tz)}_{(0,1)}$, $K^{(2tz)}_{(2,0)}$, and $K^{(2tz)}_{(0,2)}$ which all vanish identically. Finally, $K^{(2zz)}_{(2,0)}$ and $K^{(2zz)}_{(0,2)}$ are also subject to the same equation and boundary conditions and are therefore identical. These identities essentially follow from simple symmetry properties and the fact that the considered boundary metric perturbations \eqref{h1}, \eqref{h2}, \eqref{h3} do not source all 24 possible bulk fluctuations $K^{(a)}_j$.

\paragraph{Local solutions}

The $K^{(a)}_j$ have been defined to be analytic at the horizon where the local solution thus depends on a single near-horizon mode $Z^{(a)}_j$:
\begin{align}
	K^{(a)}_j(u)=Z^{(a)}_j+\sum\limits_{s\geq1}\l_{j,s}^{(a)}\,\le(1-u\ri)^s\;. \label{KNH}
\end{align}
The equations of motion \eqref{H1tEOM}, \eqref{H1zEOM}, \eqref{H2ttEOM}, \eqref{H2zzEOM}, \eqref{H2tzEOM} also show that solutions can be expanded near the boundary as
\begin{align}
	K_j^{(a)}=X_j^{(a)}+k_{j,1}^{(a)}\,u+Y_j^{(a)}\,u^2+\sum\limits_{s\geq3}k_{j,s}^{(a)}\,u^s+\log u\sum\limits_{s\geq2}l_{j,s}^{(a)}\,u^2\;, \label{KNB}
\end{align}
with leading mode $X_j^{(a)}$ and sub-leading mode $Y_j^{(a)}$. The boundary conditions discussed around eqs.~\eqref{H1t1z} and \eqref{H2ttzztz} amount to
\begin{align}
	X_0^{(1\a)}&=1\;,\;\;X_{j\geq1}^{(1\a)}=0\;,&\a\in\le\{t,z\ri\}\;,\nn\\
	X_j^{(2\b)}&=0\;,&\b\in\le\{tt,zz,tz\ri\}\;.\label{KNBcdt}
\end{align}

\paragraph{Global solutions}

Some of the $K^{(a)}_j(u)$ can be solved for analytically. Using the constraint \eqref{uConstraint} on the background, $\pa_uK^{(1t)}_0$ and $\pa_uK^{(2tt)}_{(0,0)}$ are found to satisfy homogeneous linear first-order equations. The unique regular solutions that are normalised to $1$ and $0$ at the boundary respectively are
\begin{align}
	K^{(1t)}_0=1\;,&&K^{(2tt)}_{(0,0)}=0\;. \label{pIdent1}
\end{align}
This in turn renders the equations satisfied by $K^{(2tt)}_{(1,0)}$ and $K^{(2tt)}_{(2,0)}$ homogeneous. The unique regular solutions that vanish at the boundary are identically zero,
\begin{align}
	K^{(2tt)}_{(1,0)}&=0\;, \label{etaIdent1}\\
	K^{(2tt)}_{(2,0)}&=0\;.
\end{align}
Likewise, employing the constraint \eqref{uConstraint} and replacing $A'(u)$ using eq.~\eqref{uEOM3} one can successively solve for $K^{(1z)}_0$, $K^{(2zz)}_{(0,0)}$, and $K^{(1t)}_1$:
\begin{align}
	K^{(1z)}_0&=\frac{f(u)}{1-u}\;,\;\;\;K^{(2zz)}_{(0,0)}=1-f(u)\;,\label{pIdent2}\\
	K^{(1t)}_1&=-\frac{i}{4\pi T}\log\le(\frac{f(u)}{1-u}\ri)\;. \label{etaIdent2}
\end{align}
Finally, comparing the corresponding equations of motion reveals that
\begin{align}
	K^{(2zz)}_{(2,0)}=-\le(1-u\ri)K^{(1z)}_2+K^{(2tz)}_{(1,1)}\;. \label{Kreln}
\end{align}
This leaves us with five of the initial 24 functions $K^{(a)}_j$ still undetermined:
\begin{align}
	K^{(1z)}_2\;,\;\;K^{(1t)}_2\;,\;\;K^{(2tt)}_{(1,1)}\;,\;\;K^{(2zz)}_{(1,1)}\;,\;\;K^{(2tz)}_{(1,1)}\;.\label{Kundetermined}
\end{align}
We did not manage to solve for $K^{(1z)}_2$, but we can do a little better with the other four. Owing to the residual gauge symmetry at second order $\mathcal{O}(\e^2)$ in metric perturbations, their equations of motion only depend on the functions' derivatives and take the form
\begin{align}
	\frac{\dif}{\dif u}\le[u\,f(u)\,e^{4A(u)}\,\frac{\dif}{\dif u}K^{(a)}_j(u)\ri]=u\,f(u)\,e^{4A(u)}\,\Upsilon^{(a)}_j(u)\;,\label{K1stOrder}
\end{align}
where
\begin{align}
	K^{(a)}_j\in\le\{K^{(1t)}_2,\;K^{(2tt)}_{(1,1)},\;K^{(2zz)}_{(1,1)},\;K^{(2tz)}_{(1,1)}\ri\}\;.
\end{align}
The explicit expressions for the $\Upsilon^{(a)}_j$, which only depend on the background, are written in eq.~\eqref{Ypsilons} in appendix \ref{subleadingModes}. Using that the four $K^{(a)}_j$ at hand are zero at the boundary and that the regularity condition \eqref{KNH} implies that the square bracket in eq.~\eqref{K1stOrder} vanishes at the horizon, we can formally integrate eq.~\eqref{K1stOrder} to
\begin{align}
	K^{(a)}_j(u)=\int\limits_0^u\dif v\,\frac{1}{v\,f(v)\,e^{4A(v)}}\int\limits_1^v\dif w\,w\,f(w)\,e^{4A(w)}\,\Upsilon^{(a)}_j(w)\;.\label{KintegralSoln}
\end{align}
Expanding the solution near the boundary (see appendix \ref{subleadingModes} for details) we can read off the corresponding sub-leading modes,
\begin{align}
	Y^{(a)}_j=&\frac{1}{8\pi^2T^2}\le(\frac{f_H^2\,e^{2A_H}}{4A_b}\ri)\le\{\pm\le(1-\frac{\f_L^2}{8}\ri)\ri.\nn\\
	&\le.-\int\limits_0^1\dif w\le[4w\,f(w)\,e^{4A(w)}\,\frac{\Upsilon^{(a)}_j(w)}{A_b\,L^2}\mp\frac{1}{w^2}\le(1-\frac{\f_L^2}{12}w\ri)\ri]\ri\}\;,\label{Yintegral}
\end{align}
where the upper signs refer to $K^{(a)}_j\in\le\{K^{(2tt)}_{(1,1)},\;K^{(2zz)}_{(1,1)},\;K^{(2tz)}_{(1,1)}\ri\}$ and the lower signs to $K^{(a)}_j=K^{(1t)}_2$. In the conformal case $\f=0$, eq.~\eqref{blackBranes}, the integrals can be performed analytically, resulting in
\begin{align}
	\le(Y^{(1t)}_2,\;Y^{(2tt)}_{(1,1)},\;Y^{(2zz)}_{(1,1)},\;Y^{(2tz)}_{(1,1)}\ri)\xrightarrow{\f\rightarrow0}\le(\frac{-5+4\log2}{32\pi^2T^2},\;\frac{1-\log2}{8\pi^2T^2},\;0,\;\frac{1}{8\pi^2T^2}\ri)\;.\label{YCFT}
\end{align}

\section{Analytic results for second-order transport}\label{StressTensor}

This section contains our analytic results for second-order transport in the class of non-conformal holographic models introduced in section \ref{model}. We provide explicit formulae for the five second-order coefficients \eqref{coeffs} in subsection \eqref{FormulaeCoefficients}. They apply to all holographic RG flows triggered by a scalar operator of dimension $\D=3$, at any value of the temperature. Notably, we find that the particular combination $\tilde{H}\equiv2\eta\tau_\pi-2\le(\kappa-\kappa^*\ri)-\l_2$ vanishes identically in this class of models. In subsection \eqref{proof} we prove that the identity $H=2\eta\tau_\pi-4\l_1-\l_2=0$, which is universally satisfied by infinitely strongly coupled holographic fluids with conformal symmetry \cite{Haack:2008xx}, still holds when taking into account leading non-conformal corrections to the transport coefficients.

\subsection{Formulae for transport coefficients}\label{FormulaeCoefficients}

To compute the transport coefficients we need to match the effective hydro result for the field-theory stress tensor $\le<T^{\m\n}\ri>$ with the corresponding holographic result. After a suitable renormalisation procedure, the latter can be read off from the near-boundary expansion of the dual bulk metric \cite{Myers:1999psa,deHaro:2000vlm}. Details on this calculation can be found in appendix \ref{renormalisation}. Once we apply the solutions for the bulk metric perturbations from the previous section, the result will depend on the near-boundary modes $\le\{A_b,f_b,\f_L,\f_{SL}\ri\}$ of the background, on the temperature $T$ which sets the unit of momenta, and on the sub-leading modes $\le\{Y^{(1z)}_2,Y^{(1t)}_2,Y^{(2tt)}_{(1,1)},Y^{(2zz)}_{(1,1)},Y^{(2tz)}_{(1,1)}\ri\}$ of the metric perturbations \eqref{Kundetermined} without closed-form solutions. We can somewhat simplify the result by making use of the fact that $f'(u)$ satisfies the first-order equation \eqref{uEOM3}. Imposing the requirement of asymptotically $AdS_5$, eq.~\eqref{NBu}, it can be integrated to give
\begin{align}
	f'(u)=2f_b\,\frac{A_b^2\,e^{-4A(u)}}{u}\;. \label{fprimeSoln}
\end{align}
Evaluating this at the horizon and using expressions \eqref{Hawking} for temperature $T$ and entropy density $s$ reveals that
\begin{align}
	A_b^2=-\le(\frac{4\pi G_NL}{f_b}\ri)s\,T\;.\label{Abreln}
\end{align}
Finally, we can re-express the leading- and sub-leading modes $\f_L$ and $\f_{SL}$ of the scalar in terms of source $\Lambda$ and expectation value $\le<O\ri>$ of the scalar operator, employing relations \eqref{scalarSource} and \eqref{scalarVEV} from the holographic renormalisation.

We are ready at last to present the holographic result for $\le<T^{\m\n}\ri>$ in units of the field-theory quantities $T$, $s$, $\Lambda$, and $\le<O\ri>$. The background stress tensor at $\mathcal{O}(\e^0)$ takes the ideal-fluid form
\begin{align}
	\bar{T}^{\m\n}=\begin{pmatrix} \bar{\e}&&&\\&\bar{p}&&\\&&\bar{p}&\\&&&\bar{p}\end{pmatrix} \label{bckgrdST}
\end{align}
with energy density
\begin{align}
	\bar{\e}=\frac{3}{4}\,sT-\frac{1}{4}\Lambda\le<O\ri>
\end{align}
and pressure\footnote{In the conformal case $\f=0$, eq.~\eqref{blackBranes}, this reduces to $\bar{\e}=3\bar{p}=\frac{3\pi^3L^3}{16 G_N}T^4$ or $\bar{\e}=3\bar{p}=\frac{3\pi^2}{8}N^2T^4$ for $\mathcal{N}=4$ specifically \cite{Gubser:1996de}.}
\begin{align}
	\bar{p}=\frac{1}{4}\,sT+\frac{1}{4}\Lambda\le<O\ri>\;.\label{pressure}
\end{align}
For each of the perturbations \eqref{h1}, \eqref{h2}, \eqref{h3}, the leading response of the transverse tensor component $\le<T^{xy}\ri>$ occurs at order $\mathcal{O}(\e^2)$ and indeed takes the expected hydro form \eqref{Txy1}, \eqref{Txy2}, \eqref{Txy3}. Owing to relation \eqref{Abreln}, the shear viscosity assumes its universal value in units of $s$ \cite{Kovtun:2003wp,Buchel:2003tz,Kovtun:2004de,Buchel:2004qq,Son:2007vk,Brustein:2008cg,Iqbal:2008by,Cremonini:2011iq}
\begin{align}
	\eta=\frac{1}{4\pi}s\;,\label{eta}
\end{align}
while the second-order coefficients are given by the following expressions:
\begin{subequations}
\label{2ndOrderExpr}
\begin{align}
	\kappa&=-\frac{2}{f_b}\,Y^{(2tz)}_{(1,1)}\,sT\;,\\
	\eta\,\tau_\pi+\kappa^*&=\frac{1}{f_b}\le(\frac{1}{32\pi^2T^2}+Y^{(1t)}_2-Y^{(2tz)}_{(1,1)}\ri)\,sT\;,\\
	\l_1+\frac{\kappa^*}{2}&=\frac{1}{f_b}\le(\frac{1}{32\pi^2T^2}+Y^{(1t)}_2-Y^{(2tz)}_{(1,1)}+Y^{(2tt)}_{(1,1)}\ri)\,sT\;,\\
	\l_2&=\frac{2}{f_b}\le(\frac{1}{32\pi^2T^2}+Y^{(1t)}_2+Y^{(2tz)}_{(1,1)}\ri)\,sT\;,\\
	\l_3-2\kappa^*&=\frac{4}{f_b}\,Y^{(2zz)}_{(1,1)}\,sT\;.
\end{align}
\end{subequations}
The solutions for the $Y^{(a)}_j$ that appear in eq.~\eqref{2ndOrderExpr} are stated in eq.~\eqref{Yintegral}. The fact that $\le<T^{xy}\ri>$ assumes the form dictated by hydrodynamics is a non-trivial check on our computations. In particular, the correct result for $\eta$ relies on identities \eqref{etaIdent1} and \eqref{etaIdent2}, while identities \eqref{pIdent1} and \eqref{pIdent2} are essential to ensuring that the second-order responses \eqref{Txy1} and \eqref{Txy2} both reproduce the same pressure as the background stress tensor \eqref{bckgrdST} does.

We further note that the dependence of the second-order coefficients \eqref{2ndOrderExpr} on $Y^{(1z)}_2$ cancels out as a consequence of relation \eqref{Kreln}. The \emph{five} a-priori independent combinations \eqref{2ndOrderExpr} hence only depend on the \emph{four} sub-leading modes $\le\{Y^{(1t)}_2,Y^{(2tt)}_{(1,1)},Y^{(2zz)}_{(1,1)},Y^{(2tz)}_{(1,1)}\ri\}$ given by eq.~\eqref{Yintegral}. In particular, there exists one linear combination of second-order coefficients that is independent of the $Y^{(a)}_j$:
\begin{align}
	\tilde{H}\equiv2\eta\,\tau_\pi-2\le(\kappa-\kappa^*\ri)-\l_2=0\;.\label{identity}
\end{align}
This identity does not depend on a particular background solution and therefore holds for all holographic RG flows triggered by a scalar operator of dimension $\D=3$, at any value of the temperature. We would also like to emphasise that it crucially relies on the \emph{global} solutions \eqref{pIdent1}--\eqref{Kreln} we found for some of the metric perturbations. Accordingly, identity \eqref{identity} cannot simply follow from Ward identities which rely only on the \emph{local} near-boundary solutions \cite{Henningson:1998gx,deHaro:2000vlm}. To our knowledge, the only holographic theories in which all transport coefficients entering $\tilde{H}$ have been computed are planar $\mathcal{N}=4$ (at infinite 't Hooft coupling \cite{Policastro:2001yc,Baier:2007ix,Bhattacharyya:2008jc} as well as including leading finite coupling corrections \cite{Buchel:2004di,Benincasa:2005qc,Buchel:2008ac,Buchel:2008sh,Myers:2008yi,Buchel:2008bz,Grozdanov:2014kva}) and the non-conformal Chamblin-Reall background \cite{Bigazzi:2010ku}. In both cases $\tilde{H}$ vanishes in the infinite coupling limit, but it becomes non-zero when taking into account finite coupling corrections in $\mathcal{N}=4$.

We conclude this subsection by illustrating the usability of eqs.~\eqref{2ndOrderExpr}. They allow for the straightforward computation of the second-order coefficients \eqref{coeffs} for any given background solution $A(u)$, $f(u)$, $\f(u)$ of Einstein's equations \eqref{uEOM} coupled to a scalar with potential $V$, eq.~\eqref{potential}: simply extract horizon and boundary modes of the background according to eqs.~\eqref{NHu} and \eqref{NBu}, perform integrals \eqref{Yintegral} over the background to compute the four $Y^{(a)}_j$, and plug the result into eqs.~\eqref{2ndOrderExpr}. For instance, the values \eqref{YCFT} of the $Y^{(a)}_j$ in the conformal case, eq.~\eqref{blackBranes}, readily reproduce the known results for all second-order transport coefficients in conformal holographic fluids ($\kappa^*=0$) \cite{Baier:2007ix,Bhattacharyya:2008jc},
\begin{align}
	\le\{\;\kappa,\;\eta\tau_\pi,\;\l_1,\;\l_2,\;\l_3\;\ri\}=\le(\frac{s}{8\pi^2T}\ri)\le\{\;2,\;2-\log2,\;1,\;-2\log 2,\;0\;\ri\}\;,\label{cftvalues}
\end{align}
where for $\mathcal{N}=4$ specifically $s/T=\pi^2N^2T^2/2$.


\subsection{Proof that $H=0$ to leading order away from conformality}\label{proof}

We will now prove that, to leading order in the deviation from conformality, $H=2\eta\tau_\pi-4\l_1-\l_2$ vanishes for any uncharged holographic $CFT_4$ deformed by a relevant scalar operator of dimension $\D=3$. From eq.~\eqref{2ndOrderExpr}, $H$ takes the form
\begin{align}
	H=-\frac{4}{f_b}\le(\frac{1}{32\pi^2T^2}+Y^{(1t)}_2+Y^{(2tt)}_{(1,1)}\ri)\label{deltaH}
\end{align}
in such theories, where $Y^{(1t)}_2$ and $Y^{(2tt)}_{(1,1)}$ are given by expression \eqref{Yintegral}, combined with eq.~\eqref{Ypsilons}. A priori, $Y^{(1t)}_2$ and $Y^{(2tt)}_{(1,1)}$ depend on both background fields $A(u)$, $f(u)$ and their boundary and horizon modes $\le\{A_b,f_b\ri\}$ and $\le\{A_H,f_H\ri\}$. Replacing $A(u)$ with its analytic solution, eq.~\eqref{fprimeSoln}, 
\begin{align}
	A(u)=\frac{1}{4}\log\le(\frac{2f_b\,A_b^2}{u\,f'(u)}\ri)\;,
\end{align}
however, also cancels the explicit dependence on $A_b$ and $A_H$ in eq.~\eqref{deltaH} and $H$ becomes a function of $f_b$ and $f(u)$ only, though it depends on the latter through a complicated integral over a rational functional of $f(u)$ and its derivatives. Yet, close to the UV-fixed point we can expand the integrand in the deviation $\d f(u)$ from the conformal solution \eqref{blackBranes},
\begin{subequations}
\begin{align}
	f(u)&=1-u^2+\d f(u)\;,\\
	f(u\rightarrow0)&\sim 1+\le(-1+\d f_b\ri)u^2\;,\;\;\;f(u\rightarrow1)\sim\le(2+\d f_H\ri)\le(1-u\ri)\;. \label{deltaflocal}
\end{align}
\end{subequations}
To linear order in $\d f$, the result for $H$ is
\begin{align}
	H=\frac{1}{2\pi^2 T^2}&\le(-\frac{1}{4 f_b}+\frac{1}{4}+\ri. \nn\\
	&\quad+\le.\int\limits_0^1\dif w\le[P(u)\d f''(u)+Q(u) \d f'(u)+\le(Q'(u)-P''(u)\ri)\d f(u)\ri]\ri)\;,
\end{align}
where we defined
\begin{align}
	P(u)&\equiv\frac{\le(1+u\ri)\log\le(1+u\ri)}{8u^2}\;,\\
	Q(u)&\equiv\frac{u\le(1+2u-3u^2\ri)-\le(1+u\ri)^2\le(2-3u\ri)\log\le(1+u\ri)}{8u^3\le(1-u^2\ri)}\;.
\end{align}
Integrating by parts and inserting the near-boundary and near-horizon behaviour of $\d f(u)$, eq.~\eqref{deltaflocal}, one finds that $H$ indeed vanishes to first order in $\d f$. This proves that, even when taking into account the leading non-conformal corrections to second-order transport caused by an arbitrary scalar operator of dimension $\D=3$, the combination $H= 2\eta\tau_\pi-4\l_1-\l_2$ remains zero in strongly coupled holographic fluids.


\section{Numerical results for second-order transport}\label{num}

This section contains our numerical results for second-order transport in non-conformal holographic liquids. In subsection \ref{leadingCorrection} we present the leading non-conformal corrections to second-order hydro coefficients. They only depend on the mass term in the scalar potential and are therefore common to all holographic RG flows triggered by a scalar operator of dimension $\D=3$. In subsection \ref{numFlows} we introduce two specific examples of holographic RG-flow families. We plot and discuss our numerical results for the transport coefficients along these flows in subsection \ref{plots}. In subsection \ref{entropyConstraints} we exploit known relations between transport coefficients, which must hold if the local entropy production is to be positive, in order to extend our numerical results to seven second-order coefficients.


\subsection{Leading non-conformal correction to second-order coefficients}\label{leadingCorrection}

All holographic RG-flows described by \eqref{action} and \eqref{potential} share the same UV fixed point, dual to the $AdS_5$-black brane geometry \eqref{blackBranes} with vanishing scalar $\f=0$. At high temperatures (compared to the scalar source $\Lambda$), $\f$ remains close to zero and can be treated as a small perturbation of the conformal background. Its leading backreaction on the geometry occurs at quadratic order $\mathcal{O}(\f^2)$ and can be computed analytically as we show in appendix \ref{backreaction}. The result only depends on the quadratic mass term in the bulk potential $V(\f)$ and is therefore the same for all holographic RG-flows triggered by a scalar operator of dimension $\D=3$. 

Taking the result for the backreaction (eqs.~\eqref{deltaphiModes}, \eqref{Aconvenient}, \eqref{deltaAf}, \eqref{deltaAprime}, and \eqref{deltaf}) and plugging it into integrals \eqref{Yintegral} to compute the deviation of the $Y^{(a)}_j$ from their conformal values \eqref{YCFT}, we obtain the leading non-conformal corrections to the transport coefficients via eq.~\eqref{2ndOrderExpr}. We were not able to perform the required integrals over the backreaction analytically, but they are easily evaluated numerically.

Thanks to identities $\tilde{H}=2\eta\tau_\pi-2\le(\kappa-\kappa^*\ri)-\l_2=0$ and $H=2\eta\tau_\pi-4\l_1-\l_2=0$, which we proved to hold when taking into account leading non-conformal corrections in subsection \eqref{proof}, only three of the five transport coefficients \eqref{coeffs} are independent. The results for $\kappa$,  $\l_2$, and $\l_1+\l_3/4$ are\footnote{Note that $\l_3$, as opposed to $\kappa^*$, does not vanish for conformal fluids in general \cite{Moore:2010bu,Saremi:2011nh}. It does, however, vanish in conformal holographic theories at strictly infinite coupling \cite{Bhattacharyya:2008jc}.}
\begin{subequations}\label{coeffsPerturb}
\begin{align}
	\kappa&= 2\le(\frac{s}{8\pi^2T}\ri)\le(1-4.5979\cdot10^{-3}\le(\Lambda/T\ri)^2\ri)+\mathcal{O}\le(\le(\Lambda/T\ri)^4\ri)\\
	\l_2&= -2\log2\le(\frac{s}{8\pi^2T}\ri)\le(1+4.9253\cdot10^{-3}\le(\Lambda/T\ri)^2\ri)+\mathcal{O}\le(\le(\Lambda/T\ri)^4\ri)\\
	\l_1+\l_3/4&= \le(\frac{s}{8\pi^2T}\ri)\le(1+2.5506\cdot10^{-3}\le(\Lambda/T\ri)^2\ri)+\mathcal{O}\le(\le(\Lambda/T\ri)^4\ri)\;,
\end{align}
\end{subequations}
while the other two combinations of transport coefficients satisfy
\begin{align}
	\eta\tau_\pi+\kappa^*=\kappa+\l_2/2\;,&&\l_1+\kappa^*/2=\kappa/2\;.\label{coeffsDependent}
\end{align}
The numerical integration over the backreaction can be done to very high accuracy, but we chose to only display the first five digits in eq.~\eqref{coeffsPerturb}. By checking the identity $H=0$ and by comparing results obtained when writing cancelling divergences in the integrands in different ways, we could estimate the absolute numerical error to be smaller than $10^{-14}$.

We conclude this subsection by emphasising again that the leading non-conformal corrections \eqref{coeffsPerturb} and \eqref{coeffsDependent} are common to all holographic RG-flows triggered by a scalar operator of dimension $\Delta=3$.

\subsection{Two simple families of holographic RG flows}\label{numFlows}

\paragraph{Potential $V_{(1)}$:}

The first family $V_{(1)}$ of potentials that we are going to investigate has recently been introduced in ref.~\cite{Attems:2016ugt}. It derives from a family of quartic superpotentials $W$
\begin{align}
	L\,W=-\frac{3}{2}-\frac{\f^2}{8}+\frac{\f^4}{16\f_m^2}\;,
\end{align}
resulting in
\begin{align}
	V_{(1)}&=8\le[\le(\frac{\pa W}{\pa\f}\ri)^2-\frac{2}{3}W^2\ri]\nn\\
	&=\frac{1}{L^2}\le[-12-\frac{3}{2}\f^2-\frac{1}{12}\f^4+\frac{6+\f_m^2}{12\f_m^4}\f^6-\frac{1}{48\f_m^4}\f^8\ri]\;.\label{V1}
\end{align}
The potential $V_{(1)}$ has a maximum at $\f=0$ and a minimum at the free parameter $\f_m$. Close to $\f=\f_m$, the potential takes the form
\begin{align}
	L^2\,V_{(1)}=-12\frac{L^2}{L_\mathrm{IR}^2}+\frac{m_\mathrm{IR}^2L^2}{2}\le(\f-\f_m\ri)^2+\mathcal{O}\le(\le(\f-\f_m\ri)^3\ri)\;,
\end{align}
yielding a second asymptotically $AdS_5$-region, dual to an IR fixed point, with a smaller AdS radius,
\begin{align}
	L_\mathrm{IR}\equiv\le(1+\frac{\f_m^2}{24}\ri)^{-1}L\;,
\end{align}
and a positive mass,
\begin{align}
	m_\mathrm{IR}^2L^2=12+\frac{\f_m^2}{3}\;.
\end{align}
The potential $V_{(1)}$ thus represents a family of RG flows between a UV CFT, deformed by a relevant operator of dimension $\D=3$, and an IR CFT, whose number of degrees of freedom is smaller by a factor of $\le(L_\mathrm{IR}/L\ri)^{3/2}$ compared to the UV \cite{Henningson:1998gx,Khavaev:1998fb} and which is deformed by an irrelevant operator of dimension \cite{Gubser:1998bc,Witten:1998qj}
\begin{align}
	\D_\mathrm{IR}&=2+2\sqrt{1+\frac{m_\mathrm{IR}^2L_\mathrm{IR}^2}{4}}\nn\\
	&=4+\frac{48}{24+\f_m^2}\in\le(4,6\ri)\;.\label{deltaIR}
\end{align}
For smaller $\f_m$, the number of degrees of freedom in the IR increases and the operator becomes more irrelevant in the IR. In this sense, the RG flow happens more quickly. In the opposite limit $\f_m\rightarrow\infty$, the potential becomes quartic,
\begin{align}
	V_{(1)}\xrightarrow{\f_m\rightarrow\infty}\frac{1}{L^2}\le[-12-\frac{3}{2}\f^2-\frac{1}{12}\f^4\ri]\;,\label{V1quartic}
\end{align}
the number of degrees of freedom in the IR goes to zero, and the IR operator becomes marginally irrelevant. In this sense, the RG flow happens infinitely slowly.

\begin{figure}[p!]
	\includegraphics[width=\textwidth]{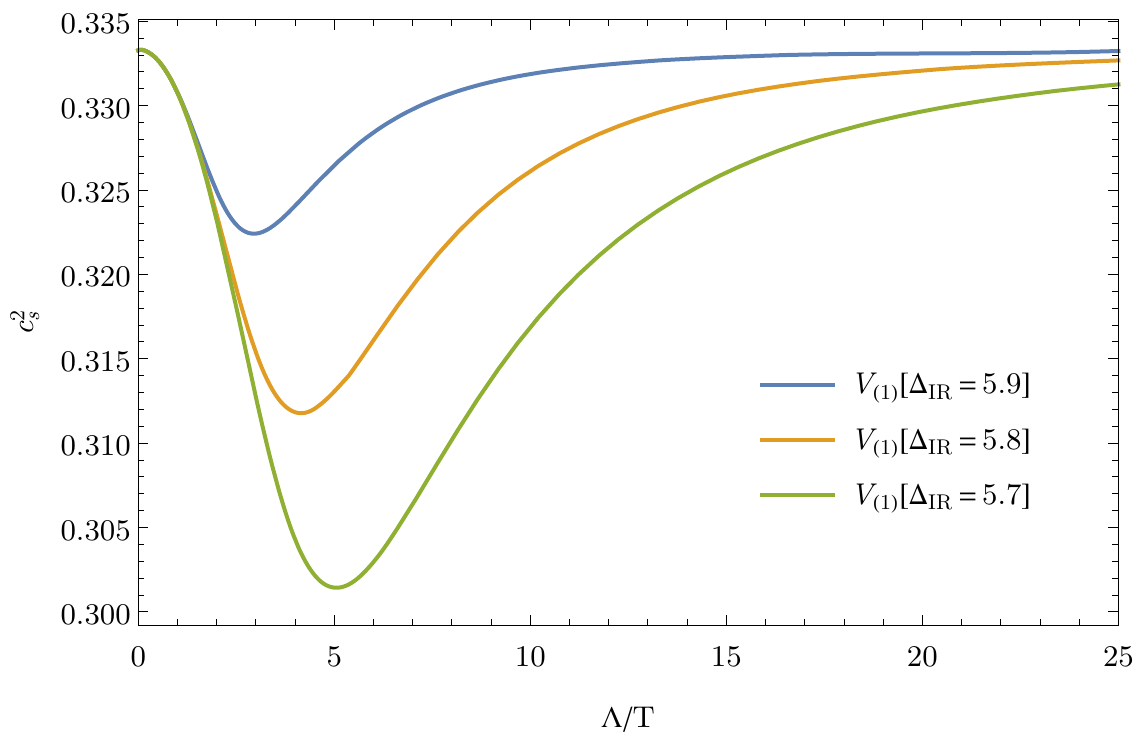}
	
	\vspace{10pt}
	
	\includegraphics[width=\textwidth]{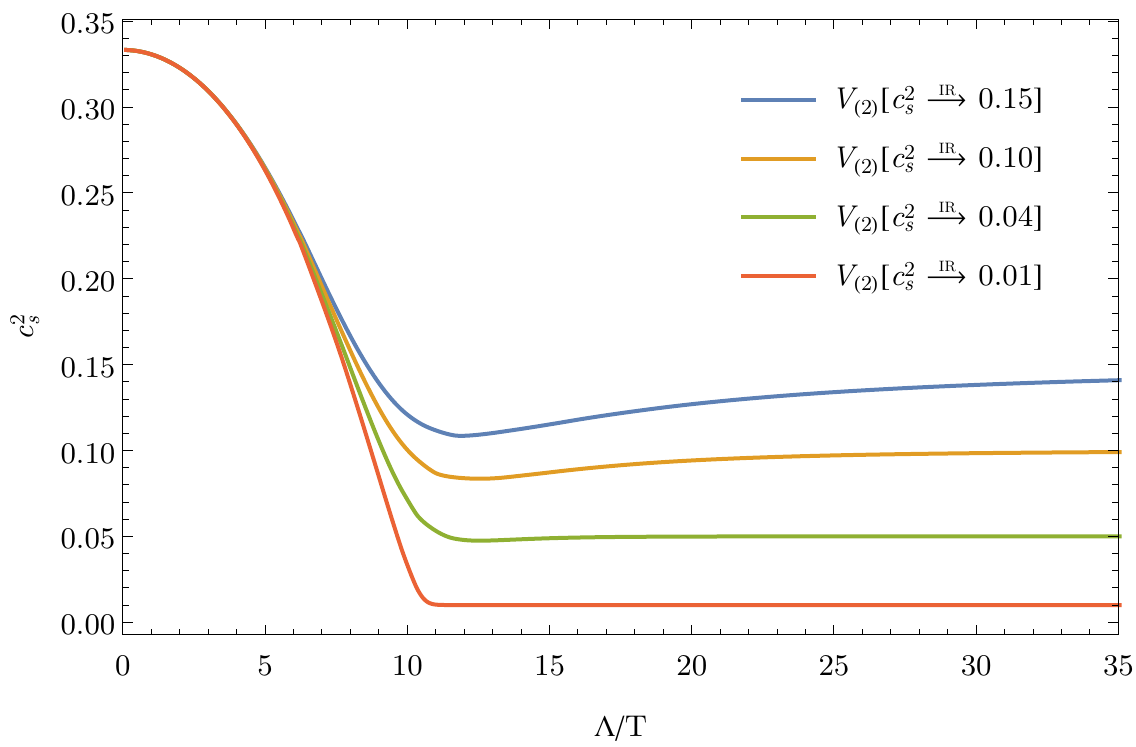}
	\caption{The speed of sound squared $c_s^2$ plotted versus operator source $\Lambda$ over temperature $T$. The upper plot shows our numerical results for potential $V_{(1)}$ for three operator dimensions $\D_\mathrm{IR}$ at the IR fixed point, eq.~\eqref{deltaIR}. The lower plot shows our numerical results for potential $V_{(2)}$ for four values of the parameter $\g$, which determines the value of $c_s^2$ in the deep IR via eq.~\eqref{cslimit}, $c_s^2\rightarrow1/3-\g^2/2$. Each curve represents a holographic RG flow triggered by a different operator.}
	\label{figure1}
\end{figure}
\afterpage{\clearpage}

\paragraph{Potential $V_{(2)}$:}
The second family $V_{(2)}$ of potentials that we are going to study is given by
\begin{align}
	V_{(2)}&=\frac{1}{L^2}\le[-12-\le(\frac{3}{2}-\frac{1}{\g^2}\ri)\f^2+\frac{2}{\g^4}\le(1-\cosh(\g\f)\ri)\ri]\label{V2}\\
	&=\frac{1}{L^2}\le[-12-\frac{3}{2}\f^2-\frac{1}{12}\f^4-\frac{\g^2}{360}\f^6+\mathcal{O}(\f^8)\ri]\;.\nn
\end{align}
The potential $V_{(2)}$ is monotonically decreasing for any value of the free parameter $\g$ and hence represents an RG flow from a UV CFT towards a non-conformal IR.

For large $\f$, $V_{(2)}$ asymptotically approaches an exponential potential, $L^2\,V_{(2)}\rightarrow -e^{\g\f}/\g^4$, for which the finite-temperature solution is given by the analytically known Chamblin-Reall background \cite{Chamblin:1999ya,Gubser:2008ny}. In the deep IR, i.e.~for large values $\f_H$ of the scalar at the horizon, the near-horizon region of solutions to our model $V_{(2)}$ is therefore asymptotically described by the Chamblin-Reall background. In particular, temperature $T$ and entropy density $s$ take the following form in the limit of large $\f_H$ \cite{Gubser:2008ny}:
\begin{subequations}
\begin{align}
	\log\le(LT\ri)&=\le(\frac{\g}{2}-\frac{1}{3\g}\ri)\f_H+\le(\text{const in $\f_H$}\ri)\;,\\
	\log \le(4G_Ns\ri)&=-\frac{\f_H}{\g}+\le(\text{const in $\f_H$}\ri)\;.
\end{align}
\end{subequations}
This implies that the speed of sound $c_s$ in the deep IR is
\begin{align}
	c_s^2=\frac{\dif\bar{p}}{\dif\bar{\e}}=\frac{\dif\log T}{\dif\log s}\xrightarrow{T\rightarrow0}\frac{1}{3}-\frac{\g^2}{2}\;.\label{cslimit}
\end{align}
Importantly, black brane solutions to $V_{(2)}$ can be stable for arbitrarily small temperatures only if $c_s^2>0$, i.e.~if $\le|\g\ri|<\sqrt{2/3}$.\footnote{Note that this implies that the top-down \emph{GPPZ}-flow with superpotential $W=-\frac{3}{4}\le(1+\cosh(\f/\sqrt{3})\ri)$ does not admit stable black-brane solutions below a certain minimum temperature as its potential approaches $V\rightarrow -\le(3/8\ri)\exp(2\f/\sqrt{3})$ for large $\f$.} Note that for $\g\rightarrow 0$, $V_{(2)}$ approaches the same quartic potential as $V_{(1)}$ does in the limit $\f_m\rightarrow\infty$, eq.~\eqref{V1quartic}:
\begin{align}
	V_{(2)}\xrightarrow{\g\rightarrow0}\frac{1}{L^2}\le[-12-\frac{3}{2}\f^2-\frac{1}{12}\f^4\ri]\;.\label{V2quartic}
\end{align}

\begin{figure}[!p]
	\includegraphics[width=\textwidth]{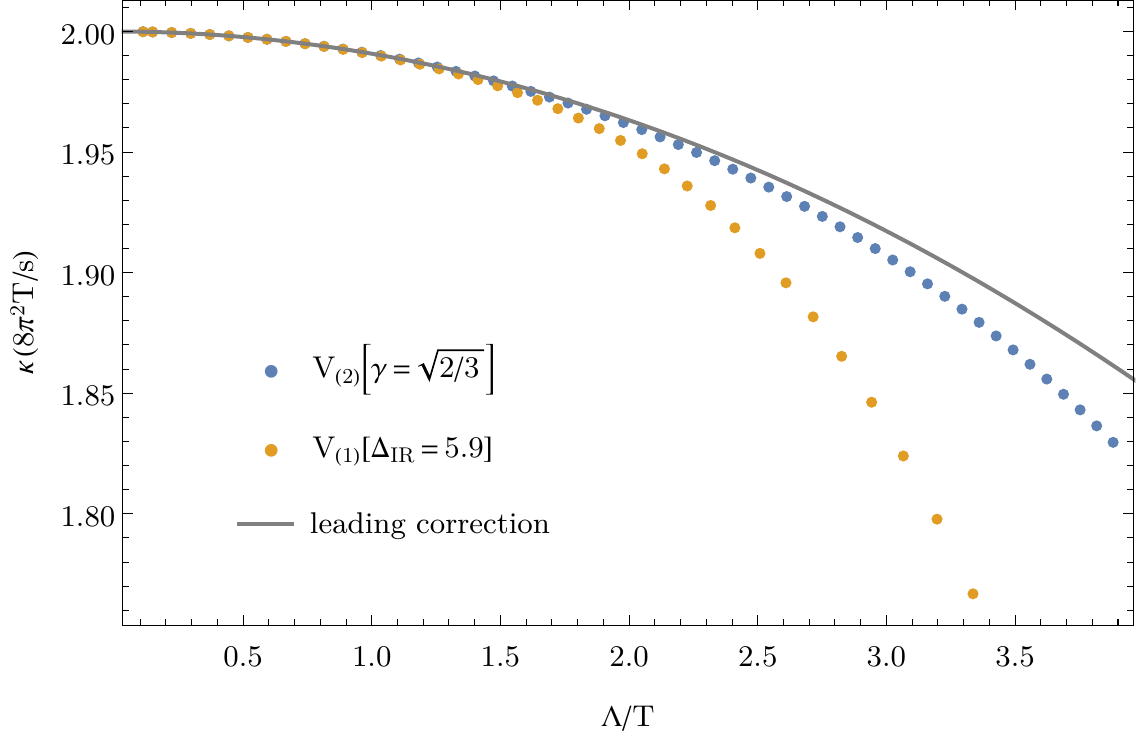}
		
	\vspace{10pt}
	
	\includegraphics[width=\textwidth]{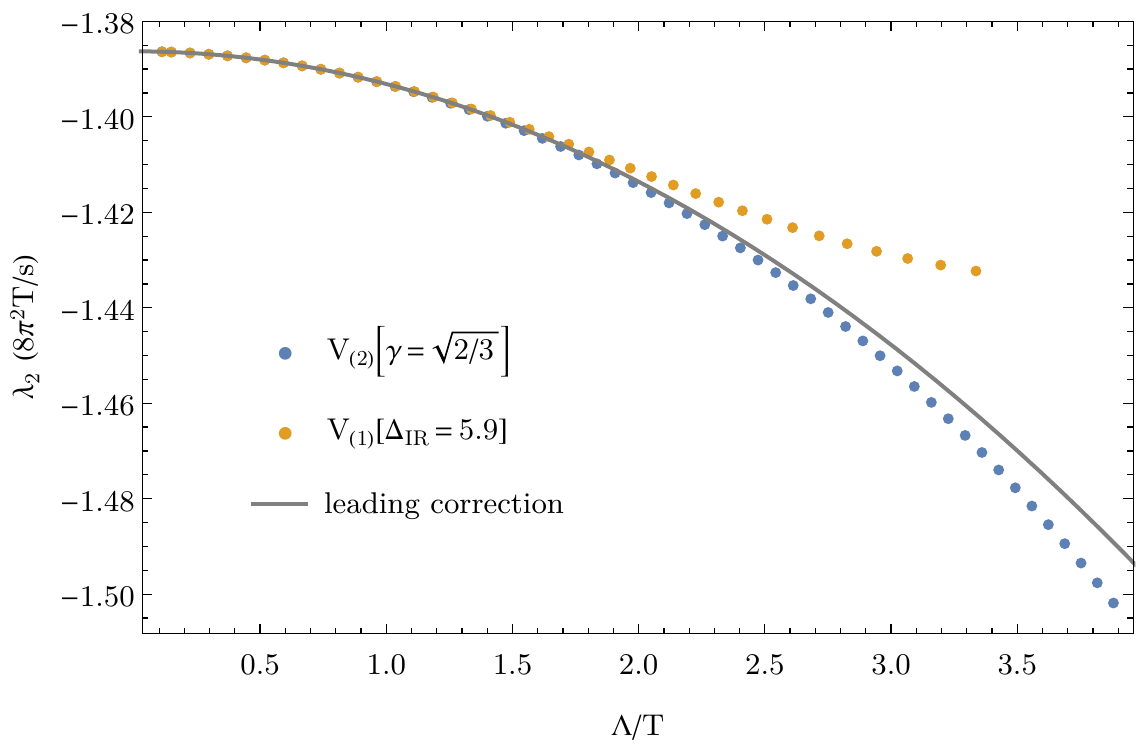}
	\caption{The second-order coefficients $\kappa$ and $\l_2$ in units of $s/\le(8\pi^2T^2\ri)$ plotted versus $\Lambda/T$. The plots show our numerical results for $V_{(1)}$ and $V_{(2)}$ with $\D_\mathrm{IR}=5.9$ and $\g=\sqrt{2/3}$ respectively. For smaller values of $\D_\mathrm{IR}$ and $\g$, the orange and blue curve move closer to each other until they coincide for $\D_\mathrm{IR}\rightarrow4$ and $\g=0$, see eqs.~\eqref{V1quartic} and \eqref{V2quartic}. The solid line describes the leading non-conformal corrections \eqref{coeffsPerturb} from subsection \ref{leadingCorrection}.}
	\label{figure2}
\end{figure}
\begin{figure}[!p]
	\includegraphics[width=\textwidth]{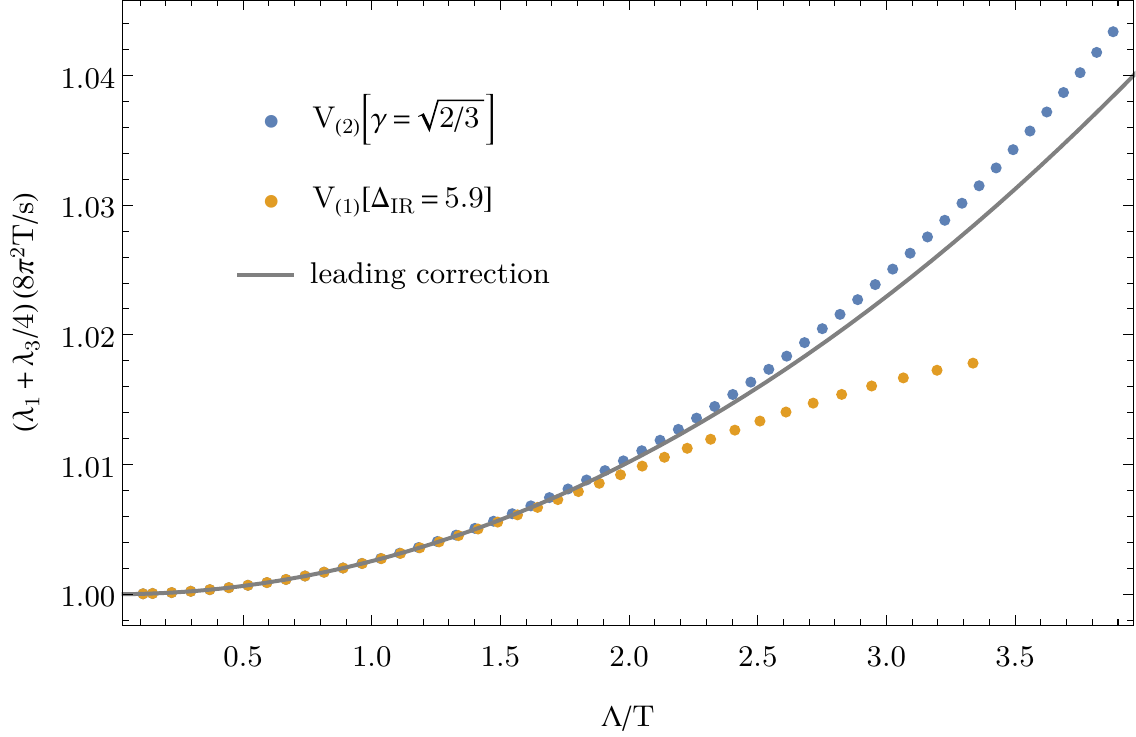}
		
	\vspace{15pt}
	
	\includegraphics[width=\textwidth]{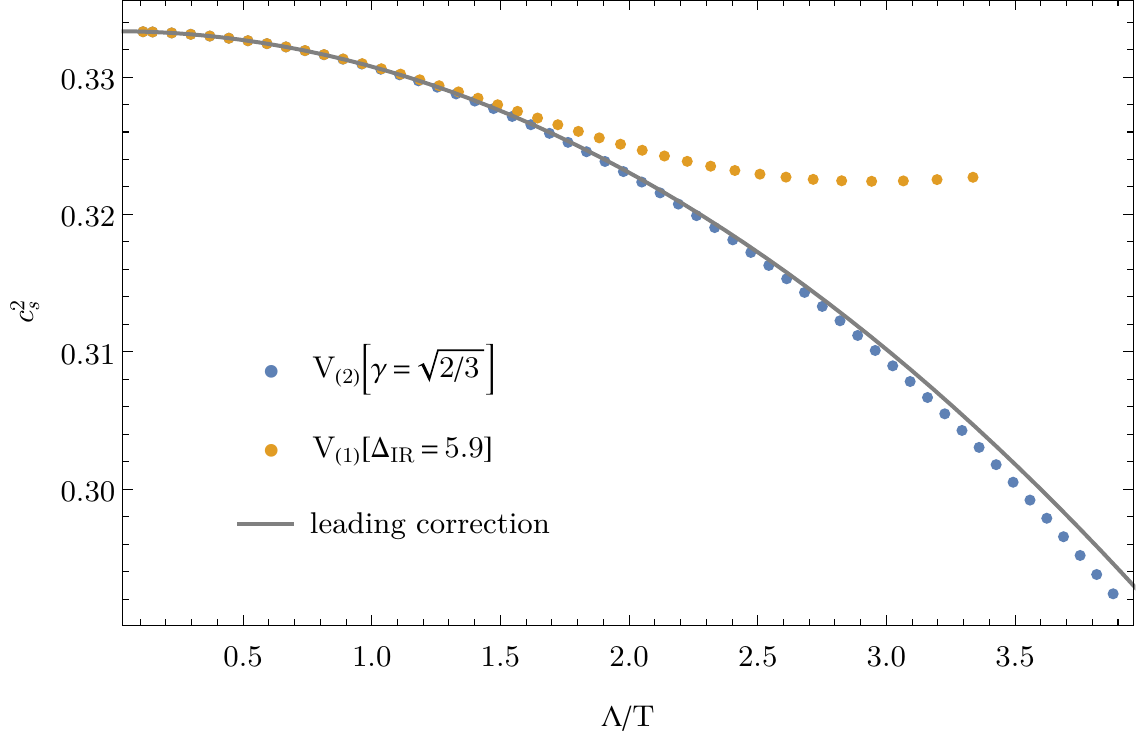}
	\caption{The second-order coefficient $\l_1+\l_3/4$ in units of $s/\le(8\pi^2T^2\ri)$ and the speed of sound squared $c_s^2$ plotted versus $\Lambda/T$. The plots show our numerical results for $V_{(1)}$ and $V_{(2)}$ with $\D_\mathrm{IR}=5.9$ and $\g=\sqrt{2/3}$ respectively. For smaller values of $\D_\mathrm{IR}$ and $\g$, the orange and blue curve move closer to each other until they coincide for $\D_\mathrm{IR}\rightarrow4$ and $\g=0$, see eqs.~\eqref{V1quartic} and \eqref{V2quartic}. The solid line describes the leading non-conformal corrections \eqref{coeffsPerturb} from subsection \ref{leadingCorrection}.}
	\label{figure3}
\end{figure}

\afterpage{\clearpage}

\paragraph{Background solutions}
For both families $V_{(1)}$ and $V_{(2)}$ we used the method devised in ref.~\cite{Gubser:2008ny} to construct numerical background solutions. The essential steps are summarised in appendix \ref{numericalConstr}. Figure \ref{figure1} shows our numerical results for the speed of sound along the two RG flows for a few representative values of the respective parameters $\D_\mathrm{IR}$ and $\g$.


\subsection{Second-order coefficients along examples of RG flows}\label{plots}

This section contains our numerical results for the second-order transport coefficients \eqref{coeffs} along the two families of holographic RG flows introduced in subsection \ref{numFlows}. For each of the two families $V_{(1)}$ and $V_{(2)}$, we looked at around 20 parameter values covering the range $4.1\leq\D_\mathrm{IR}\leq5.9$ and $0\leq\g\leq\sqrt{2/3}$ respectively. For each of these flows we then constructed numerical background solutions at about 40 different temperatures and computed the second-order coefficients from eq.~\eqref{2ndOrderExpr}.

Our main result is that the combination $H=2\eta\tau_\pi-4\l_1-\l_2$ vanishes in all cases considered, even when the individual transport coefficients deviate from their conformal values by factors of two and more. More precisely, the absolute values we obtained for $H$ all lie below our numerical accuracy of order $10^{-5}$. For details on the numerics see the end of appendix \ref{numericalConstr}. Our result suggests that the identity $H=0$ does not only hold for holographic fluids with conformal symmetry \cite{Haack:2008xx,Shaverin:2012kv,Grozdanov:2015asa,Shaverin:2015vda,Grozdanov:2014kva} or close to a fixed point (see ref.~\cite{Bigazzi:2010ku} and our subsection \ref{proof}), but is in fact universally satisfied by all holographic fluids at infinite coupling, with or without conformal symmetry.

Combined with $\tilde{H}=2\eta\tau_\pi-2\le(\kappa-\kappa^*\ri)-\l_2=0$, the result $H=0$ implies that only three of the five coefficients \eqref{coeffs} are independent. In figures \ref{figure2} and \ref{figure3} we plot the second-order coefficients $\kappa$, $\l_2$, $\l_1+\l_3/4$ and the speed of sound squared $c_s^2$ versus $\Lambda/T$. The plots show our numerical results for $V_{(1)}$ and $V_{(2)}$, each with the largest parameter value considered, i.e.~$\D_\mathrm{IR}=5.9$ and $\g=\sqrt{2/3}$. Our results for smaller parameter values all lie between these two extreme curves and vary smoothly with $\D_\mathrm{IR}$ and $\g$. The plots confirm that the behaviour of the transport coefficients close to the UV fixed point, i.e.~for small $\Lambda/T$, is well described by the leading non-conformal correction discussed in subsection \ref{leadingCorrection}. If we compare figures \ref{figure2} and \ref{figure3} with figure \ref{figure1}, the difference in the considered range of $\le(\Lambda/T\ri)$-values stands out. In particular, while figure \ref{figure1} follows the speed of sound all the way from the UV to the IR region, figures \ref{figure2} and \ref{figure3} only contain results for relatively high $T$ and do not capture the IR properties of $V_{(1)}$ and $V_{(2)}$. This is due to the fact that we did not use the same radial coordinate to compute the transport coefficients that we used to compute thermodynamic quantities such as $c_s^2$. The latter were obtained using the scalar $\f$ itself as radial coordinate, as required by the method we employed to construct background solutions \cite{Gubser:2008ny}, see appendix \ref{numericalConstr} for details. However, $\f$ is not a suitable coordinate in the UV, where it becomes small everywhere. In particular, it does not lend itself to a perturbative treatment as in subsection \ref{leadingCorrection} and it is ill-defined in the conformal limit $\f\rightarrow0$. For this reason, we switched to the $u$-coordinate when dealing with metric fluctuations around the background. While $u$ is well-defined in the UV, it becomes problematic in the IR because the region $\f\in\le(0,\f_H\ri)$ is mapped onto the same interval $u\in\le(0,1\ri)$ for all values of $\f_H$. At low temperatures, i.e.~for large $\f_H$, the modes in the $u$-coordinate become very large and render the numerics unstable. Nonetheless, we decided to work in the $u$-coordinate as it allowed us to obtain independent results from the perturbative treatment of $\f$ and to compare every step of our calculations with the conformal case. The drawback is that reliable results for the transport coefficients could only be obtained for relatively small values of $\Lambda/T$. In particular, we cannot observe how the transport coefficients go back to their conformal values in the case of $V_{(1)}$ or begin to approach the values assumed in the Chamblin-Reall background in the case of $V_{(2)}$ (see appendix \ref{ChamblinReall}). We leave the numerical investigation of second-order coefficients in the deep IR for future research.

Let us take another look at figure \ref{figure1}. It indicates that for $V_{(2)}$ the influence of the IR becomes dominant only if $\Lambda/T\gtrsim10$. We found that the same is true for $V_{(1)}$ with $\D_\mathrm{IR}\lesssim5.2$. In these cases it was therefore possible to obtain reliable numerical results for larger values of $\Lambda/T$ than it was in the case of $V_{(1)}$ with $\D_\mathrm{IR}$ close to 6. Figure \ref{figure4} shows the deviations of $c_s^2$, $\kappa$, $\l_2$, and $\l_1+\l_3/4$ from their conformal values for $V_{(2)}$ with $\g=\sqrt{2/3}$ and $\g=0$, plotted against $\Lambda/T$. Results for $V_{(2)}$ with $0<\g<\sqrt{2/3}$ lie between these two curves. Results for $V_{(1)}$ with $\D_\mathrm{IR}\lesssim5.2$ closely follow the curve for $V_{(2)}|_{\g=0}$, in agreement with $V_{(1)}\xrightarrow{\D_\mathrm{IR}\rightarrow4}V_{(2)}|_{\g=0}$ from eqs.~\eqref{V1quartic} and \eqref{V2quartic}.


\subsection{Employing relations from the entropy current}\label{entropyConstraints}

\begingroup
\allowdisplaybreaks
Patches of local equilibrium in a fluid obey the second law of thermodynamics \cite{Kovtun:2012rj}. This requires the existence of an entropy current whose divergence is non-negative when the equations of motion are satisfied. A sufficient condition, which ensures that the second law is obeyed, is to demand that at each order in the entropy current's gradient expansion only terms that always lead to non-negative entropy production can appear. Imposing this condition, refs.~\cite{Romatschke:2009kr,Bhattacharyya:2012nq} found five equalities relating second-order coefficients. The same relations were found in ref.~\cite{Jensen:2012jh} by coupling the fluid to external sources. Written in our conventions these equalities can be found in ref.~\cite{Moore:2012tc}. In particular, they determine $\kappa^*$, $\xi_5$, and $\xi_3+\xi_6$ in terms of $\kappa(T)$ and $\l_3(T)$ as
\begin{figure}[t!b]
	\includegraphics[width=\textwidth]{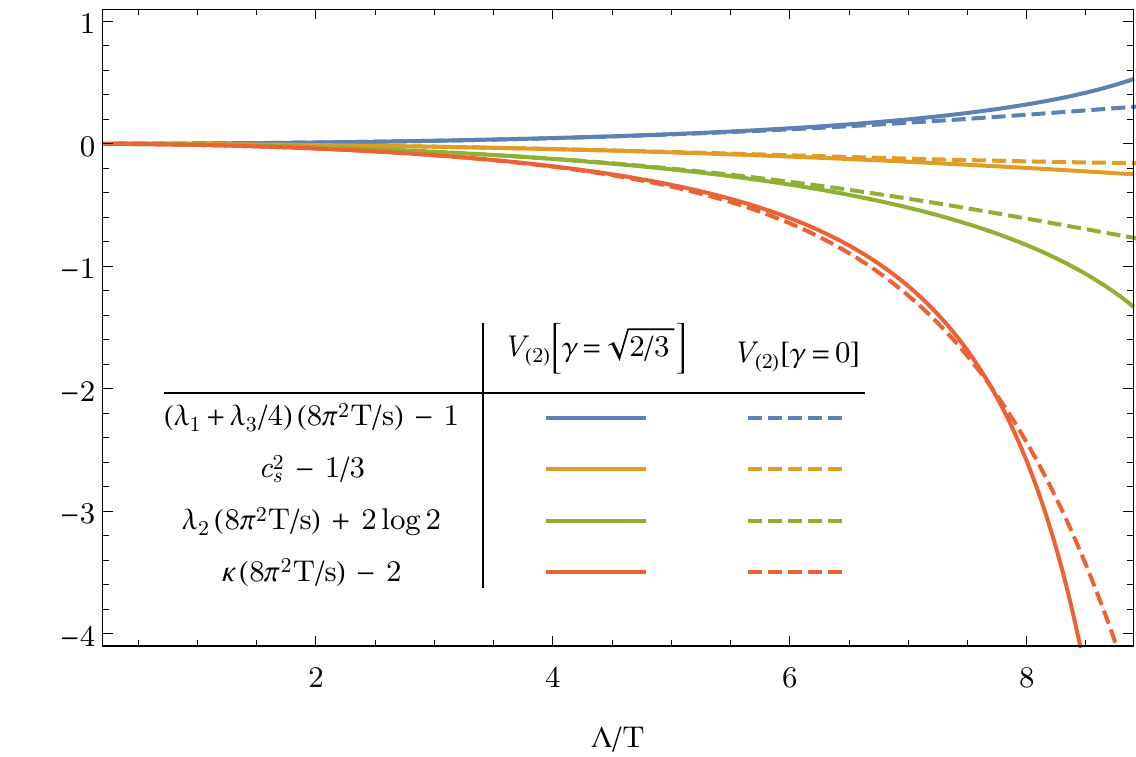}
	\caption{The deviation of the speed of sound squared $c_s^2$ and the second-order coefficients $\l_1+\l_3/4$, $\l_2$, $\kappa$ in units of $s/\le(8\pi^2T\ri)$ from their conformal values \eqref{cftvalues}, versus $\Lambda/T$. The solid line shows our numerical results for $V_{(2)}$ with $\g=\sqrt{2/3}$, the dashed line refers to $V_2$ with $\g=0$. Results for intermediate values of $\g$ interpolate smoothly between the two lines.}
	\label{figure4}
\end{figure}
\begin{align}
	\kappa^* =&\; \kappa-\frac{T}{2}\frac{d\kappa}{dT}\;,\hspace{3cm} \xi_5 = \frac{1}{2}\le(c_s^2T\frac{d\kappa}{dT}-c_s^2\kappa-\frac{\kappa}{3}\ri)\;,\\
	\xi_3 + \xi_6 =&\; \le[\frac{1}{3}\le(1-3c_s^2\ri)+\frac{T}{12}\le(1-6c_s^2\ri)\frac{d}{dT}+\frac{T^2 c_s^2}{4}\frac{d^2}{dT^2}\ri]\kappa+\le[\frac{1}{12}\le(1-9c_s^2\ri)+\frac{T c_s^2}{4}\frac{d}{dT}\ri]\l_3\nn\;.
\end{align}
Making use of these relations we can extend our results to all five conformal coefficients $\kappa$, $\eta\tau_\pi$, $\l_1$, $\l_2$, $\l_3$ and to three non-conformal coefficients $\kappa^*$, $\xi_5$, $\xi_3+\xi_6$. They are shown in figure \ref{figure5}.
\endgroup

\begin{figure}[!p]
	\includegraphics[width=\textwidth]{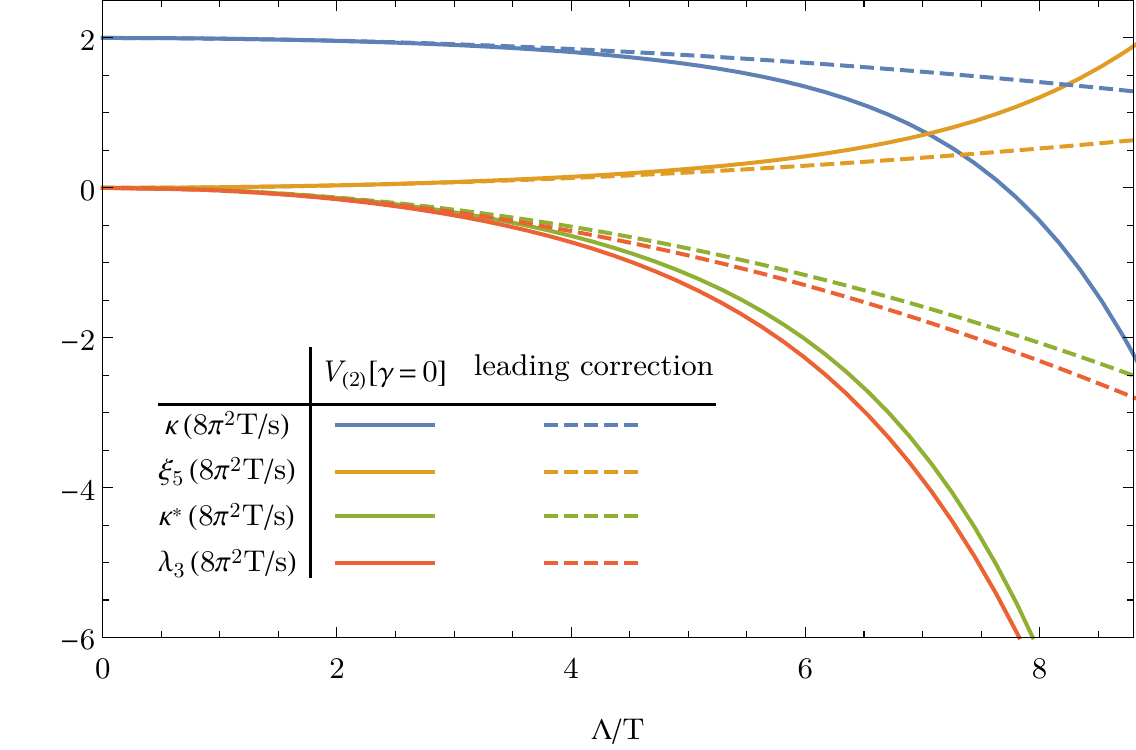}
		
	\vspace{10pt}
	
	\includegraphics[width=\textwidth]{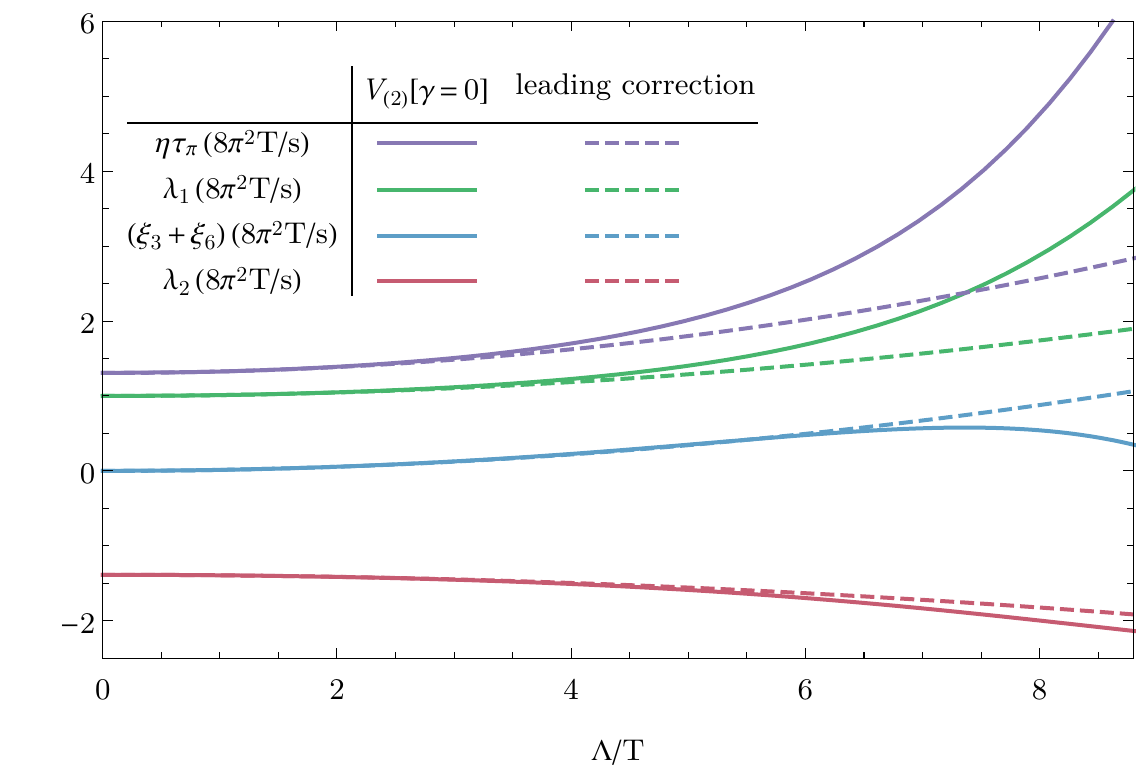}
	\caption{All five conformal second-order coefficients $\kappa$, $\eta\tau_\pi$, $\l_1$, $\l_2$, $\l_3$ and three non-conformal second-order coefficients $\kappa^*$, $\xi_5$, $\xi_3+\xi_6$ in units of $s/\le(8\pi^2T\ri)$ versus $\Lambda/T$. The solid line shows our numerical results for $V_{(2)}$ with $\g=0$, the dashed line describes the leading non-conformal corrections from subsection \ref{leadingCorrection}. The eight second-order coefficients were obtained applying constraints derived from the positivity of the local entropy production as described in subsection~\ref{entropyConstraints}.}
	\label{figure5}
\end{figure}

\afterpage{\clearpage}


\section{Conclusion and outlook}\label{conclusion}

In this paper we studied second-order hydrodynamic transport in strongly coupled non-conformal field theories. We derived new Kubo formulae for five second-order transport coefficients valid for uncharged non-conformal fluids in $(3+1)$ dimensions. We applied these Kubo formulae to holographic RG flows triggered by a relevant scalar operator of dimension $\D=3$ and found expressions for the five second-order transport coefficients at infinite coupling in terms of the holographically dual gravitational background solution. We showed that the relation
\begin{align}
	\tilde{H}=2\eta\tau_\pi-2\le(\kappa-\kappa^*\ri)-\l_2=0
\end{align}
is satisfied along all such holographic RG flows. We proved that the Haack-Yarom identity \cite{Erdmenger:2008rm,Haack:2008xx}
\begin{align}
	H=2\eta\tau_\pi-4\l_1-\l_2=0\;,
\end{align}
which is known to hold for conformal holographic fluids at infinite coupling, is also satisfied when leading-order non-conformal corrections are included within the class of considered RG flows. For the two specific classes of RG flows we studied numerically, we found the Haack-Yarom identity to be satisfied all along each flow beyond perturbative non-conformal corrections. This provides further evidence that the identity may be universally satisfied by strongly coupled fluids \cite{Grozdanov:2014kva,Shaverin:2012kv,Grozdanov:2015asa,Shaverin:2015vda,Bigazzi:2010ku,Wu:2016erb}.

In section~\ref{hydroSection} we derived a new set of Kubo formulae, eq.~\eqref{Kubo}, for five second-order transport coefficients of non-conformal fluids and showed how these Kubo formulae can be applied to strongly coupled fluids with a gravity dual. We introduced a specific class of strongly coupled non-conformal field theories that describe an RG flow induced in the UV by a scalar operator of dimension $\D=3$ in section~\ref{model}. We derived bulk equations of motion both for asymptotically AdS backgrounds of the dual Einstein-scalar models and for the metric fluctuations on such backgrounds that are relevant for the Kubo formulae. In section~\ref{einstein} we solved the fluctuation equations in a hydrodynamic derivative expansion and derived explicit integral expressions for the sub-leading modes in terms of background data, eq.~\eqref{Yintegral}. In section~\ref{StressTensor} we presented our analytic results for second-order transport coefficients. We determined the expressions \eqref{2ndOrderExpr} for five transport coefficients in terms of the dual background data and proved that a certain linear combination of second-order transport coefficients, $\tilde{H}=2\eta\,\tau_\pi-2\le(\kappa-\kappa^*\ri)-\l_2$, vanishes identically along any RG flow in the class considered. We showed that the Haack-Yarom identity $H=0$, eq.~\eqref{universalIdentity}, is obeyed to leading order in the deviation from conformality for arbitrary holographic RG flows triggered by a relevant scalar operator of dimension $\D=3$. In section~\ref{num} we presented our numerical results for second-order transport coefficients. First, we computed the leading non-conformal corrections to the second-order coefficients. Second, we introduced two specific families of $\D=3$ operators and numerically found the Haack-Yarom identity to be obeyed along both families of corresponding RG flows. Third, we plotted the independent combinations of second-order coefficients along both classes of RG flows and found them to agree with the perturbative results in the appropriate high-temperature regime. Fourth, we applied three constraints derived from positivity of the local entropy production to extend our results to all five conformal coefficients $\kappa$, $\eta\tau_\pi$, $\l_1$, $\l_2$, $\l_3$ and to three non-conformal coefficients $\kappa^*$, $\xi_5$, $\xi_3+\xi_6$.

Our work points to a number of open questions. For technical reasons, we restricted our bulk computations to transverse vector and tensor channel perturbations of the metric. If scalar sound perturbations were included as well and an appropriate set of Kubo formulae were derived, it would be possible to compute all fifteen second-order non-conformal transport coefficients. The major technical obstacle is that these scalar sound perturbations necessarily source fluctuations of the scalar field in the bulk. Nevertheless, if one computed all fifteen second-order coefficients one could check the five relations which have been derived from the positivity of the local entropy production \cite{Romatschke:2009kr,Bhattacharyya:2012nq,Jensen:2012jh}.

In the presentation of our numerical results in subsections~\ref{plots} and \ref{entropyConstraints} we pointed out that our choice of the radial coordinate $u$ made it impossible to numerically access the deep IR of the considered RG flows, despite several other advantages. As we explain in appendix~\ref{numericalConstr}, the IR region would become accessible if we used the scalar $\f$ as radial coordinate instead. We have precise expectations as to what such a numerical study would reveal. For the first family of potentials the transport coefficients have to return to their conformal values. For the second family of potentials they are expected to approach the values they assume in the Chamblin-Reall background, which we list in appendix~\ref{ChamblinReall}.

Whether our proof in subsection~\ref{proof} that $H$ vanishes when taking into account leading non-conformal corrections caused by an operator of dimension $\D=3$ can be generalised to relevant operators of arbitrary dimension $2<\D<4$ is another open question. Since the computation of $H$ does not involve fluctuations of the bulk scalar, a change in the dual operator dimension would require a different set of holographic counterterms but would not affect the relevant bulk metric fluctuations.

What our results entail for the entropy current is another direction for future research. Fluids with simple gravity duals at strong coupling have been conjectured to obey a principle of minimal dissipation \cite{Haehl:2014zda,Grozdanov:2014kva}. The observation that the lower bound on the shear viscosity over entropy density ratio is universally satisfied by a large class of holographic theories \cite{Kovtun:2003wp,Buchel:2003tz,Kovtun:2004de,Son:2007vk,Iqbal:2008by} is a first hint in this direction since this ratio appears as coefficient of the leading contribution to the entropy production. At second order in gradients, the entropy current of a conformal fluid contains two terms \cite{Romatschke:2009kr}: the coefficient of the first one vanishes if $2\l_1=\kappa$, which is indeed true for conformal holographic fluids at infinite coupling \cite{Baier:2007ix}, while the coefficient of the second term remains unknown. The relations $\tilde{H}=0$ and $H=0$ found in this work are equivalent to $H=0$ and $2\l_1=\kappa-\kappa^*$. It was shown within an effective action approach to adiabatic hydrodynamics that these two relations must hold for perfect fluids that do not produce entropy \cite{Haehl:2015pja}. Indeed, the Haack-Yarom identity, $H=0$, seems to require either infinite coupling or adiabaticity. Even for conformal fluids, the identity is violated in examples of weakly coupled systems in the kinetic regime \cite{York:2008rr} and when finite coupling corrections are included in the hypothetical dual of Gauss-Bonnet gravity \cite{Grozdanov:2014kva,Grozdanov:2015asa,Shaverin:2015vda}.\footnote{Consistently, $H=0$ does not follow from the generalised Onsager relations, which were derived from an effective action for hydrodynamics in ref.~\cite{Crossley:2015evo} and which should apply to any uncharged conformal fluid.} It would be interesting to explore if the relations $H=0$ and $2\l_1=\kappa-\kappa^*$ lead to cancellations in the divergence of the entropy current similar to the conformal case. This would provide further evidence in favour of the principle of minimal dissipation.

We think that the results of this paper will provide a valuable starting point for research into the open questions outlined above.

\section*{Acknowledgements}

We would like to thank A.~Starinets for suggesting this topic to us and for his invaluable feedback. We thank J.~Casalderrey-Solana, A.~Ficnar, and S.~Grozdanov for helpful discussions as well as A.~O'Bannon, S.~Grozdanov, and M.~Aspioti for valuable comments on the manuscript. This work was supported by the European Research Council under the European Union's Seventh Framework Programme (ERC Grant agreement 307955). P.~K.\ acknowledges partial support by the Science and Technology Facilities Council. J.~P.\ is supported by the Clarendon Fund and by St John's College, Oxford.

\clearpage
\appendix

\section{Second-order constitutive relations}
\label{AppendixConstituent}

This appendix contains the explicit constitutive relations for the stress tensor of an uncharged relativistic fluid up to second order in the gradient expansion that were used to derive \eqref{bdyPert} and \eqref{Txyresponse}. Firstly, let the 4-velocity be the unit timelike eigenvector of $\le<T^{\m\n}\ri>$ and $\e$ the corresponding eigenvalue to \emph{all} orders in the gradient expansion (Landau frame) \cite{LandauLifshitz,Kovtun:2012rj,Romatschke:2009kr}. Further define the projection to symmetric, traceless tensors that are transverse to the fluid motion,
\begin{align}
	\D^{\m\n}(x)&\equiv u^\m(x)u^\n(x)+g_{(0)}^{\m\n}(x)\;,\\
	A^{<\m\n>}&\equiv\frac{1}{2}\D^{\m\r}\le(A_{\r\s}+A_{\s\r}\ri)\D^{\s\n}-\frac{1}{3}\D^{\m\n}\le(\D^{\s\r}A_{\s\r}\ri)\;,
\end{align}
the shear tensor
\begin{align}
	\s^{\m\n}\equiv2\nabla^{<\m}u^{\n>}\;,
\end{align}
and the vorticity tensor
\begin{align}
	\Omega^{\m\n}\equiv\frac{1}{2}\D^{\m\r}\le(\nabla_\r u_\s-\nabla_\s u_\r\ri)\D^{\s\n}\;.
\end{align}
The constitutive relation for $\le<T^{\m\n}\ri>$ can then be written as
\begin{align}
	\le<T^{\m\n}(x)\ri>=\e(x)u^\m(x)u^\n(x)+p(x)\D^{\m\n}(x)+\Pi^{\m\n}_\mathrm{conf.}(x)+\Pi^{\m\n}_\mathrm{non-conf.}(x)+\mathcal{O}(\pa^3)\label{constituent2}
\end{align}
where
\begin{align}
	\Pi^{\m\n}_\mathrm{conf.}(x)&\equiv-\eta\,\s^{\m\n}\nn\\
	&\quad+\eta\,\tau_\pi \le[\le(u^\l\nabla_\l\s\ri)^{<\m\n>}+\frac{1}{3}\s^{\m\n}\le(\nabla\cdot u\ri)\ri]\nn\\
	&\quad+\kappa\le[R^{<\m\n>}-2u_\l R^{\l<\m\n>\kappa}u_\kappa\ri]\nn\\
	&\quad+\l_1\s_\l{}^{<\m}\s^{\n>\l}+\l_2 \s_\l{}^{<\m}\Omega^{\n>\l} -\l_3\Omega_\l{}^{<\m}\Omega^{\n>\l}
\end{align}
is present in conformal and non-conformal fluids and was first derived in ref.~\cite{Baier:2007ix}, and where
\begin{align}
	\Pi^{\m\n}_\mathrm{non-conf.}(x)&\equiv-\z\,\D^{\m\n}\le(\nabla\cdot u\ri)\nn\\
	&\quad+\eta\,\tau_\pi^*\,\frac{1}{3}\s^{\m\n}\le(\nabla\cdot u\ri)+\kappa^*\,2u_\l R^{\l<\m\n>\kappa}u_\kappa+\l_4\nabla^{<\m}\log s\nabla^{\n>}\log s\nn\\
	&\quad+\le(\zeta\,\tau_\Pi\,u^\l\nabla_\l\le(\nabla\cdot u\ri)+\xi_1\,\s^{\kappa\l}\s_{\kappa\l}+\xi_2\le(\nabla\cdot u\ri)^2+\xi_3\,\Omega^{\kappa\l}\Omega_{\kappa\l}\ri.\nn\\
	&\quad+\le. \xi_4\,\D_\kappa{}^\l\le(\nabla_\l\log s\ri)\D^{\kappa\r}\le(\nabla_\r\log s\ri)+\xi_5\,R+\xi_6\,u^\kappa u^\l R_{\kappa\l}\ri)\D^{\m\n}\;,
\end{align}
was constructed in ref.~\cite{Romatschke:2009kr} and vanishes for conformal fluids.

\section{Sub-leading modes of metric perturbations}\label{subleadingModes}

The functionals $\Upsilon^{(a)}_j$ that appear in eq.~\eqref{K1stOrder} are
\begin{subequations}
\label{Ypsilons}
\begin{align}
	\Upsilon^{(2tt)}_{(1,1)}(u)&=\frac{L^2}{4f(u)^2}\le(\frac{1}{u^2\,e^{2A(u)}}-\frac{f'(u)^2}{f_H^2\,e^{2A_H}}\ri)\;,\\
	\Upsilon^{(2zz)}_{(1,1)}(u)&=\frac{L^2}{4u^2\,f(u)\,e^{2A(u)}}\le(2-f(u)\ri)\;,\\
	\Upsilon^{(2tz)}_{(1,1)}(u)&=\frac{L^2}{4u^2\,f(u)\,e^{2A(u)}}\;,\\
	\Upsilon^{(1t)}_2(u)&=-\frac{L^2}{4f(u)}\le\{\frac{1}{u^2\,f(u)\,e^{2A(u)}}\ri.\nn\\
	&\le.\quad+\frac{f+2\le(1-u\ri)f'-\log\le(\frac{1-u}{f}\ri)\le[\frac{f}{u}+4\le(1-u\ri)A'\,f+\le(1-u\ri)f'\ri]}{\le(1-u\ri)^2f_H^2\,e^{2A_H}}\ri\}\;.
\end{align}
\end{subequations}
In all four cases the near-boundary expansion of the first integrand in eq.~\eqref{KintegralSoln} reads
\begin{align}
	w\,f(w)\,e^{4A(w)}\,\Upsilon^{(a)}_j(w)=\pm\frac{A_b\,L^2}{4}\le(\frac{1}{w^2}-\frac{\f_L^2}{12}\frac{1}{w}\ri)+\mathcal{O}(w^0)\;,
\end{align}
where upper signs in this appendix refer to $a\in\le\{2tt,2zz,2tz\ri\}$, $j=(1,1)$, and lower signs refer to $a=1t$, $j=2$. The first integral in eq.~\eqref{KintegralSoln} thus admits the expansion
\begin{align}
	\int\limits_1^v\dif w\,w\,f(w)\,e^{4A(w)}\,\Upsilon^{(a)}_j=\mp\frac{A_b\,L^2}{4}\le(\frac{1}{v}+\frac{\f_L^2}{12}\log v\ri)+c^{(a)}_j+\mathcal{O}(v)\;,
\end{align}
where the $v$-independent contribution $c^{(a)}_j$ can be extracted via
\begin{align}
	c^{(a)}_j=\mp\frac{A_b\,L^2}{4}+\int\limits_1^0\dif w\le[w\,f\,e^{4A}\,\Upsilon^{(a)}_j\mp\frac{A_b\,L^2}{4w^2}\le(1-\frac{\f_L^2}{12}w\ri)\ri]\;.
\end{align}
Plugging this into eq.~\eqref{KintegralSoln}, together with
\begin{align}
	\frac{1}{v\,f(v)e^{4A(v)}}=\frac{v}{A_b^2}\le(1+\frac{\f_L^2}{6}v+\mathcal{O}(v^2)\ri)\;,
\end{align}
one finally obtains the near-boundary expansion of the four $K^{(a)}_j$,
\begin{align}
	K^{(a)}_j=\mp\frac{L^2}{4A_b}u\le(1+\frac{\f_L^2}{24}\,u\log u\ri)+\frac{1}{2A_b^2}\le(c^{(a)}_j\mp\frac{A_b\,L^2\,\f_L^2}{32}\ri)u^2+o(u^2)\;,
\end{align}
from which one can read off the sub-leading modes
\begin{align}
	Y^{(a)}_j = \frac{1}{2A_b^2}\le(c^{(a)}_j\mp\frac{A_b\,L^2\,\f_L^2}{32}\ri)\;,
\end{align}
yielding eq.~\eqref{Yintegral} (recall expression \eqref{Hawking} for the temperature).


\section{Holographic renormalisation}\label{renormalisation}

Holographic QFTs, like any QFT, contain UV-divergences that need to be regulated and then cancelled by appropriate counterterms in order to obtain finite, renormalised physical quantities \cite{Witten:1998qj,Skenderis:2002wp}. For theories with a holographic dual in asymptotically AdS these UV-divergences manifest themselves as near-boundary divergences of the dual gravity action \cite{Susskind:1998dq,Henningson:1998gx,Balasubramanian:1999re,Emparan:1999pm,deHaro:2000vlm}. For the holographic renormalisation procedure it is convenient to switch to a Fefferman-Graham gauge in which the asymptotically AdS-metric, eqs.~\eqref{uGauge} and \eqref{fullMetric}, takes the form
\begin{align}
	g_{mn}\dif x^m\dif x^n&=\frac{L^2}{4\r^2}{\dif \r^2}+\frac{L^2}{\r}\bar{g}_{\m\n}\dif x^\m\dif x^\n\;,\\
	\bar{g}_{\m\n}(\r,x)&=g_{(0)\m\n}(x)+\r\,g_{(2)\m\n}+\r^2\le[g_{(4)\m\n}+h_{(4)\m\n}\log\r\ri]+\dots\;,
\end{align}
the radial coordinates $u$ and $\r$ being related via
\begin{align}
	u=\frac{A_b}{L^2}\r+\frac{f_b}{4}\le(\frac{A_b}{L^2}\r\ri)^3+\dots\;,
\end{align}
ensuring that $g_{(0)\m\n}$ represents the field-theory metric. The bare gravity action can be regulated by introducing a near-boundary cut-off $\r=\varepsilon$. Denoting the outward-pointing unit normal of the regulating surface $\r=\varepsilon$ by $n_m$, $n=-\dif\r\,L/\le(2\varepsilon\ri)$, and its first and second fundamental form by
\begin{align}
	\g_{mn}=g_{mn}-n_mn_n\;,&&K_{mn}=\g_m{}^p\nabla_pn_n\;,
\end{align}
the regulated bare gravity action, eq.~\eqref{action}, including the Gibbons-Hawking term reads
\begin{align}
	S_\mathrm{reg}(\varepsilon)=\frac{1}{16\pi G_N}\int\limits_{\r\geq\varepsilon}\dif^5x\sqrt{-g}\le[R-\frac{1}{2}\le(\pa\f\ri)^2-V\ri]+\frac{1}{16\pi G_N}\int\limits_{\r=\varepsilon}\dif^4x\sqrt{-\g}\,2K^m{}_m\;.
\end{align}
The UV-divergences only depend on the fields' leading near-boundary modes which are fully determined by the  operator dimension $\D=3$. For the scalar potentials we consider, eq.~\eqref{potential}, the divergences are thus the same as for the \emph{GPPZ}-flow \cite{Girardello:1999bd}, which has been renormalised in refs.~\cite{Bianchi:2001de,Bianchi:2001kw}, and can be removed by adding the following local counterterms:
\begin{align}
	S_\mathrm{ct}^\mathrm{cov}&=-\frac{1}{8\pi G_N}\,\frac{1}{L}\int\limits_{\r=\varepsilon}\dif^4x\sqrt{-\g}\le(3+\frac{L^2}{4}R_{(\g)}+\frac{1}{4}\f^2\ri)\;,\\
	S_\mathrm{ct}^\mathrm{log}&=-\frac{1}{8\pi G_N}\int\limits_{\r=\varepsilon}\dif^4x\sqrt{-\g}\,\frac{1}{2}\log\le(\frac{\varepsilon}{L^2}\ri)\le[\frac{L^3}{8}\le(-R_{(\g)\m\n}R_{(\g)}^{\m\n}+\frac{1}{3}R_{(\g)}^2\ri)\ri.\nn\\
	&\le.\quad+\frac{L}{4}\le(-\f\,\Box_{(\g)}\f+\frac{1}{6}R_{(\g)}\f^2\ri)\ri]\;,
\end{align}
where $R_{(\g)\m\n}$ and $\Box_{(\g)}$ respectively denote Ricci tensor and Laplacian of the induced metric $\g_{\m\n}$ on the regulating surface. The explicit dependence of $S_\mathrm{ct}^\mathrm{log}$ on the regulator $\varepsilon$ breaks the invariance under bulk diffeomorphisms that induce Weyl transformations of the boundary metric $g_{(0)\m\n}$. This results in the correct conformal anomaly $\mathcal{A}$ in the dual field theory \cite{Henningson:1998gx,Bianchi:2001de,Bianchi:2001kw},
\begin{align}
	\mathcal{A}=\frac{4L^3}{16\pi G_N}\le[\frac{1}{16}\le(R_{(0)\m\n}R_{(0)}^{\m\n}-\frac{1}{3}R_{(0)}^2\ri)+\frac{1}{8}\le(-\le(\nabla_{(0)}\Lambda\ri)^2-\frac{1}{6}R_{(0)}\Lambda^2\ri)\ri]\;,
\end{align}
where $R_{(0)\m\n}$ is the Ricci tensor of the field-theory metric $g_{(0)\m\n}$, $\nabla_{(0)}$ is the associated covariant derivative, and $\Lambda$ is the source of the scalar operator, 
\begin{align}
	\f=\Lambda\sqrt{\r}+\dots\;,&&\Lambda=\frac{\sqrt{A_b}}{L}\f_L\;.\label{scalarSource}
\end{align}
We further choose to add the following finite counterterm, corresponding to a particular RG-scheme in which the renormalised stress tensor does not explicitly depend on the scalar source $\Lambda$\footnote{Another finite counterterm that one could add is $\mathrm{const}\cdot\int_{\r=\varepsilon}\dif^4x\sqrt{-\g}\,\f^4$. For instance, if one was interested in domain-wall solutions to a superpotential $W$ that preserve supersymmetry at $T=0$ one would have to choose $\mathrm{const}=-\frac{1}{8\pi G_N}\,\frac{2}{L}\,4!\le.\frac{\dif^4W}{\dif\f^4}\ri|_{\f=0}$. Being only interested in solutions at finite $T$ we choose to work in a "minimal-subtraction" scheme instead, setting $\mathrm{const}=0$.}:
\begin{align}
	S_\mathrm{ct}^\mathrm{finite}=\frac{1}{2}\int\dif^4x\sqrt{-g_{(0)}}\le(\log A_b+\frac{1}{2}\ri)\mathcal{A}\;.
\end{align}
The individual contributions to the gravity stress tensor $\mathcal{T}_{\m\n}$ \cite{Balasubramanian:1999re},
\begin{align}
	\mathcal{T}_{\m\n}=-\frac{2}{\sqrt{-\g}}\frac{\d}{\d\g^{\m\n}}\le(S_\mathrm{reg}+S_\mathrm{ct}^\mathrm{cov}+S_\mathrm{ct}^\mathrm{log}+S_\mathrm{ct}^\mathrm{finite}\ri)\;,
\end{align}
are then as follows:
\begin{align}
	-\frac{2}{\sqrt{-\g}}\frac{\d}{\d\g^{\m\n}}S_\mathrm{reg}&=\frac{1}{8\pi G_N}\le[K\,\g_{\m\n}-K_{\m\n}\ri]\;,\\
	-\frac{2}{\sqrt{-\g}}\frac{\d}{\d\g^{\m\n}}S_\mathrm{ct}^\mathrm{cov}&=\frac{1}{8\pi G_N}\le[-\frac{3}{L}\g_{\m\n}+\frac{L}{2}\le(R_{(\g)\m\n}-\frac{1}{2}R_{(\g)}\g_{\m\n}\ri)-\frac{1}{4L}\g_{\m\n}\f^2\ri]\;,\\
	-\frac{2}{\sqrt{-\g}}\frac{\d}{\d\g^{\m\n}}S_\mathrm{ct}^\mathrm{log}&=-\frac{1}{2}\mathcal{T}^\mathcal{A}_{\m\n}\le(\log\le(\frac{\varepsilon}{L^2}\ri)-\frac{1}{2}\ri)\;,\\
	-\frac{2}{\sqrt{-\g}}\frac{\d}{\d\g^{\m\n}}S_\mathrm{ct}^\mathrm{finite}&=-\frac{1}{2}\mathcal{T}^\mathcal{A}_{\m\n}\le(\log A_b+\frac{1}{2}\ri)\;,
\end{align}
with
\begin{align}
	\mathcal{T}^\mathcal{A}_{\m\n}\equiv-\frac{2}{\sqrt{-\g}}\frac{\d}{\d\g_{\m\n}}\int\dif^4x\sqrt{-g_{(0)}}\le(-\mathcal{A}\ri)
\end{align}
and \cite{Bianchi:2001de,Bianchi:2001kw}
\begin{align}
	\frac{16\pi G_N}{4L^3}\lim\limits_{\varepsilon\rightarrow0}\le(\frac{L^2}{\varepsilon}\frac{1}{2}\mathcal{T}^\mathcal{A}_{\m\n}\ri)=:\frac{1}{2}T^\mathcal{A}_{\m\n}=h_{(4)\m\n}-\frac{1}{2}g_{(0)\m\n}\le(\Lambda\,\Box_{(0)}\Lambda-\frac{1}{6}R_{(0)}\Lambda^2\ri)\;.
\end{align}
Inserting the near-boundary solutions \eqref{NBu}, \eqref{KNB} of background and metric perturbations, and employing the identities \eqref{pIdent1}--\eqref{Kreln} one can compute the renormalised field-theory stress tensor:
\begin{align}
	\le<T^{\m\n}\ri>&=g_{(0)}^{\m\a}g_{(0)}^{\n\b}\lim\limits_{\varepsilon\rightarrow0}\le(\le(\frac{L^2}{\varepsilon}\ri)\mathcal{T}_{\a\b}\ri)\;.
\end{align}
To zeroth order $\mathcal{O}(\epsilon^0)$ in metric perturbations, $\le<T^{\m\n}\ri>$ is given by the ideal-fluid stress tensor \eqref{bckgrdST}. The transverse tensor component $\le<T^{xy}\ri>$ can be trusted up to $\mathcal{O}(\e^2)$ as discussed in section \ref{HologComput} and is indeed found to take the hydro form \eqref{Txy1}, \eqref{Txy2}, \eqref{Txy3} for all three perturbations \eqref{case1Bulk}, \eqref{case2Bulk}, \eqref{case3Bulk}, with transport coefficients given by \eqref{pressure}, \eqref{eta} and \eqref{2ndOrderExpr}.

The renormalised scalar one-point function is given by
\begin{align}
	\le<O\ri>&=\lim\limits_{\varepsilon\rightarrow0}\le(\frac{1}{\le(\varepsilon^2/L^4\ri)\sqrt{-\g}}\frac{\d\le(S_\mathrm{reg}+S_\mathrm{ct}\ri)}{\d\f/\sqrt{\varepsilon}}\ri)\nn\\
	&=\lim\limits_{\varepsilon\rightarrow0}\le(\frac{L^4}{\varepsilon^{3/2}}\frac{1}{\sqrt{-\g}}\le(-\pi_\f+\frac{\d S_\mathrm{ct}}{\d\f/\sqrt{\varepsilon}}\ri)\ri)\;,
\end{align}
where
\begin{align}
	\pi_\f=-\frac{1}{16\pi G_N}\sqrt{-g}\,g^{\r\r}\pa_\r\f
\end{align}
denotes the scalar's canonical momentum. Inserting the near-boundary solutions \eqref{NBu}, one finds to first order in metric perturbations 
\begin{align}
	\le<O\ri>=\frac{L^3}{8\pi G_N}\le(\frac{A_b}{L^2}\ri)^{3/2}\f_{SL}+\mathcal{O}(\e^2)\;.\label{scalarVEV}
\end{align}
We checked that the conformal Ward identity,
\begin{align}
	\le<T^\m{}_\m\ri>=\le<O\ri>\,\Lambda+\mathcal{A}\;,
\end{align}
is satisfied to order $\mathcal{O}(\e)$ included as required (note that $\mathcal{A}=\mathcal{O}(\e^2)$).

\section{Leading backreaction of the scalar on AdS-black branes}\label{backreaction}

This appendix describes the computation of the scalar's leading backreaction on the background metric used in subsection \ref{leadingCorrection}. The calculation goes along the lines presented in ref.~\cite{Cherman:2009kf}. Just this once we will be keeping the operator dimension $\D$ general, $2<\D<4$, as it can be done without difficulty and the general form of the results might be useful in other contexts. We thus consider potentials of the form
\begin{align}
	V(\f)=-\frac{12}{L^2}+\frac{\D\le(\D-4\ri)}{2L^2}\f^2+\mathcal{O}(\f^3)\;.
\end{align}
At zeroth order, $\f=0$, the background equations of motion $\eqref{uEOM}$ are solved by the $AdS_5$-black brane metric \eqref{blackBranes}, dual to the UV-CFT. To first order in $\f$, the regular solution to the scalar equation of motion, eq.~\eqref{uEOM1}, linearised around the black-brane background is given by 
\begin{align}
	\f(u)=\d\f(u)&\equiv\f_H\,{}_2F_1(1-\D/4,\D/4;1;1-1/u^2)\nn\\
	&=\d\f_L\,u^{\le(4-\D\ri)/2}\,{}_2F_1(1-\D/4,1-\D/4;2-\D/2;u^2)\nn\\
	&\quad+\d\f_{SL}\,u^{\D/2}\,{}_2F_1(\D/4,\D/4;\D/2;u^2) \label{deltaPhi}
\end{align}
with near-boundary modes
\begin{align}
	\d\f_L=\f_H\frac{\Gamma(\D/2-1)}{\Gamma(\D/4)^2}\;,&&\d\f_{SL}=\f_H\frac{\tan(\pi\D/4)}{2\pi}\frac{\Gamma(\D/4)^2}{\Gamma(\D/2)}\;.\label{deltaphiModes}
\end{align}
At second order, the scalar itself remains unchanged, but it backreacts on the background metric. Generally, the scalar backreacts on the geometry at even orders while $\f$ itself receives corrections to its linearised solution \eqref{deltaPhi} at odd orders. There is a minor complication concerning the boundary conditions: full non-perturbative solutions to the background equations of motion \eqref{uEOM} depend on the single integration constant $\f_H$ which parameterises the single physical parameter $T/\Lambda$. In the perturbative solution, on the other hand, we need to pick the value of the scalar at the horizon at each order in the perturbative series. This apparent ambiguity is resolved by the requirement that a chosen physical observable remain unchanged order by order. We simply choose to hold $\f_H$ fixed, meaning that sub-leading corrections to $\f(u)$ need to vanish at the horizon. Other possible, albeit more complicated, choices include fixing $A_H$ or $T/\Lambda$.

A related subtlety is constituted by the fact that $A(u)$ only enters the equations of motion \eqref{uEOM} via its derivative $A'(u)$. This is most conveniently dealt with by separating from the full non-perturbative $A(u)=\frac{1}{2}\log\le(A_b/u\ri)+\mathcal{O}(u)$ the part $\tilde{A}(u)$ which vanishes at the boundary $u=0$, i.e.~we write
\begin{align}
	A(u)&=\frac{1}{2}\log\le(\frac{A_b}{u}\ri)+\tilde{A}(u)\nn\\
	&=\frac{1}{2}\log\le(\frac{A_b}{u}\ri)+\int\limits_0^u\dif v\,\tilde{A}'(v)\;. \label{Aconvenient}
\end{align}
The constant $A_b$ is fixed by the full global solution of the scalar as follows. In terms of the dimensionful $\z$-coordinate, eq.~\eqref{zGauge2}, close to the $AdS_5$ boundary we have $\f\sim\le(\Lambda \z\ri)^{4-\D}$ and $A\sim\log(L/\z)$, hence
\begin{align}
	A\sim\log\le(\Lambda L\f^{-1/\le(4-\D\ri)}\ri)
\end{align}
or, changing back to the $u$-coordinate,
\begin{align}
	A=\frac{1}{2}\log\le(\frac{\le(\Lambda\,L\,\f_L^{-1/\le(4-\Delta\ri)}\ri)^2}{u}\ri)+\mathcal{O}(u)\;,
\end{align}
and hence
\begin{align}
	A_b=\le(\Lambda\,L\,\f_L^{-1/\le(4-\Delta\ri)}\ri)^2\;. \label{AbphiL}
\end{align}
The value of $A(u)$, eq.~\eqref{Aconvenient}, at the horizon and the Hawking temperature, eq.~\eqref{Hawking}, are therefore given by
\begin{align}
	A_H=\log\le(\Lambda\,L\,\f_L^{-1/\le(4-\Delta\ri)}\ri)+I\;,&&T&=\Lambda\le(\frac{f_H\,e^I}{2\pi\,\f_L^{1/\le(4-\Delta\ri)}}\ri)\;, \label{AHT}
\end{align}
where we defined
\begin{align}
	I \equiv \int\limits_0^1\dif u\,\tilde{A}'(u)\;.
\end{align}
We emphasise again that these relations fix the observables $A_H$ and $T/\Lambda$ in terms of the full solutions for $\f_L$ and $\tilde{A}(u)$. In the particular case of a perturbative solution, they determine $A_H$ and $T/\Lambda$ order by order in the expansion parameter $\f_H$.

Let us now compute the leading, quadratic backreaction of the scalar on the geometry, which takes the form
\begin{align}
	\tilde{A}(u)=\d A(u)\;,&&f(u)=1-u^2+\d f(u)\;.\label{deltaAf}
\end{align}
The corrections $\d A$ and $\d f$ are both of order $\mathcal{O}(\f_H^2)$ and, from eqs.~\eqref{uEOM2} and \eqref{uConstraint}, satisfy the two independent equations
\begin{subequations}
\begin{align}
	\d A''+\frac{1}{u}\d A'&=-\frac{1}{6}\le(\d\f'\ri)^2\;,\label{deltaAeqn}\\
	\d f'-\frac{2}{u}\d f&=-4\le(2-u^2\ri)\d A'-\frac{\D\le(4-\D\ri)}{12u}\le(\d\f\ri)^2-\frac{u\le(1-u^2\ri)}{3}\le(\d\f'\ri)^2\;.\label{deltafeqn}
\end{align}
\end{subequations}
The second-order equation \eqref{uEOM3} for $f$ is redundant as it follows from the remaining three equations \eqref{uEOM1}, \eqref{uEOM2} and \eqref{uConstraint}. Demanding that $\d A$ vanish at the boundary $u=0$ we can integrate eq.~\eqref{deltaAeqn} twice and obtain
\begin{align}
	\d A(u)&=-\int\limits_0^v\dif v\,\frac{1}{v}\int\limits_0^v\dif w\,w\frac{1}{6}\le(\d\f'(v)\ri)^2\label{deltaAprime}\\
	&\xrightarrow{u\rightarrow0}-\frac{\f_L^2}{24}\,u^{4-\D}\;,\nn
\end{align}
in accordance with eq.~\eqref{ANB}. Requiring that $\d f$ vanish at the horizon $u=1$ we can now integrate eq.~\eqref{deltafeqn} to get
\begin{align}
	\d f(u)=-u^2\int\limits_1^u\dif v\,\frac{1}{v^2} \le[4\le(2-v^2\ri)\d A'(v) +\frac{\D\le(4-\D\ri)}{12v}\le(\d\f(v)\ri)^2 \ri.\nn\\
	\le.+\frac{v\le(1-v^2\ri)}{3}\le(\d\f'(v)\ri)^2\ri]\label{deltaf}
\end{align}
The corresponding change in $f_H=-f'(u=1)$ can be read off from eq.~\eqref{deltafeqn}:
\begin{align}
	\d f_H&=\frac{\D\le(4-\D\ri)}{12}\f_H^2+4\,\d A'(u=1)\\
	&=\frac{\D\le(4-\D\ri)}{12}\f_H^2\le(1-\frac{\D\le(4-\D\ri)}{8}\int\limits_0^1\dif v\,v^{-5}\le[{}_2F_1(2-\D/4,1+\D/4;2;1-1/v^2)\ri]^2\ri) \;.\nn
\end{align}
Defining $\d I\equiv\int_0^1\dif u\,\d A'(u)$, we obtain the following expressions for temperature $T$ and entropy density $s$ from \eqref{AHT}:
\begin{subequations}\label{sTpert}
\begin{align}
	4 G_Ns&=e^{3A_H}=\frac{\le(\Lambda\,L\ri)^3}{\f_L^{3/\le(4-\Delta\ri)}}\le(1+3\d I+\mathcal{O}(\f_H^4)\ri)\;,\\
	T/\Lambda&=\frac{\le(2+\d f_H\ri)\le(1+\d I\ri)+\mathcal{O}(\f_H^4)} {2\pi\f_L^{1/\le(4-\Delta\ri)}}\;.
\end{align}
\end{subequations}
Note that sub-leading corrections to the scalar,
\begin{align}
	\f_L=\d\f_L+\mathcal{O}(\f_H^3)=\frac{\Gamma(\Delta/2-1)}{\Gamma(\Delta/4)^2}\f_H\le(1+\mathcal{O}(\f_H^2)\ri)\;,
\end{align}
enter expressions \eqref{sTpert} at the same order as $\d f_H$ and $\d I$ do, so
\begin{align}
	\le(\frac{\Lambda}{\pi T}\ri)^{4-\D}=\frac{\Gamma(\D/2-1)}{\Gamma(\D/4)^2}\f_H\le(1+\mathcal{O}(\f_H^2)\ri)\;.\label{Tpert}
\end{align}
The sub-leading corrections to $\f_L$ cancel, however, in the expression for the speed of sound, which becomes\footnote{The result for $c_s^2$ agrees with the one computed in ref.~\cite{Cherman:2009kf} using a different radial coordinate. Note that the leading correction to the conformal value $1/3$ is negative for all $\D$, $2<\D<4$.}
\begin{align}
	c_s^2=\frac{\dif\bar{p}}{\dif\bar{\e}}=\frac{\dif \log T}{\dif \log s}=\frac{1}{3}\le[1-\le(4-\D\ri)\d f_H\ri]+\mathcal{O}\le(\f_H^4\ri)\;.
\end{align}


\section{Numerical construction of RG-flow geometries}\label{numericalConstr}

In this appendix we outline our construction of numerical background solutions to the action \eqref{action} for the scalar potentials $V_{(1)}$ and $V_{(2)}$ discussed in section \ref{num}. For that purpose it is convenient to use the scalar $\f$ as radial coordinate \cite{Gubser:2008ny}, which is assumed to increase monotonically from the boundary $\f=0$ to the horizon $\f=\f_H>0$. Defining
\begin{align}
	e^{B(\f)}\equiv\frac{L}{2u}\frac{\dif u}{\dif \f}\;,\label{uphiReln}
\end{align}
the background metric \eqref{uGauge} takes the form
\begin{align}
	\dif s^2=g_{mn}^{(0)}\dif x^m\dif x^n=e^{2A(\f)}\le[-f(\f)\dif t^2+\dif\underline{x}^2\ri]+\frac{e^{2B(\f)}}{f(\f)}\dif \f^2\;.\label{phiGauge}
\end{align}
The residual scaling symmetry, inherited from the UV CFT, can be used to fix the value of the scalar source to $\Lambda =1/L$ as the temperature is varied, knowing that all observables can only depend on the dimensionless ratio $T/\Lambda$. 

The equations of motion \eqref{uEOM} become
\begin{subequations}
\label{phiEOM}
\begin{align}
	4\frac{\dif A}{\dif\f}-\frac{\dif B}{\dif\f}+\frac{1}{f}\le(\frac{\dif f}{\dif\f}\ri)-\frac{e^{2B}}{f}\le(\frac{\dif V}{\dif\f}\ri)&=0\;,\label{phiEOM1}\\
	\frac{\dif^2A}{\dif\f^2}-\le(\frac{\dif A}{\dif\f}\ri) \le(\frac{\dif B}{\dif\f}\ri) +\frac{1}{6}&=0\;,\label{phiEOM2}\\
	\frac{\dif^2f}{\dif\f^2}+\le[4\frac{\dif A}{\dif\f}-\frac{\dif B}{\dif\f}\ri]\frac{\dif f}{\dif\f}&=0\;,\label{phiEOM3}\\
	6\le(\frac{\dif A}{\dif\f}\ri) \le(\frac{\dif f}{\dif\f}\ri) +f\le[24\le(\frac{\dif A}{\dif\f}\ri)^2-1\ri]+2e^{2B}V&=0\;,\label{phiConstraint}
\end{align}
\end{subequations}
composed of a first-order equation for $B$ \eqref{phiEOM1}, two second-order equations \eqref{phiEOM2}--\eqref{phiEOM3} for $A$ and $f$, and the first-order constraint \eqref{phiConstraint}. The system is partly redundant in the sense that the constraint \eqref{phiConstraint} and its derivative are algebraically given in terms of the other three equations:
\begin{align}
	\le(\frac{\dif}{\dif\phi}-2\frac{\dif B}{\dif \f}\ri)\eqref{phiConstraint}=-2f\eqref{phiEOM1}+\le(48 f\frac{\dif A}{\dif \phi}+6\frac{\dif f}{\dif\phi}\ri)\eqref{phiEOM2}+6\frac{\dif A}{\dif\phi}\eqref{phiEOM3} \label{phiRedundancy}
\end{align}
This redundancy prevents the constraint from restricting the series coefficients in the local near-horizon and near-boundary solutions which thus each involve five integration constants. 

Imposing regularity on $A$ and $B$, the near-horizon expansion depends on three modes $\le\{A_H,f_H^\f,\f_H\ri\}$ and reads
\begin{subequations}
\label{NHphi}
\begin{align}
	A(\f)&=A_H-\frac{1}{3}\frac{V(\f_H)}{V'(\f_H)}\le(\f-\f_H\ri)+\sum\limits_{k\geq2}b_k^{A,\f}\le(\f-\f_H\ri)^k\;,\label{ANH}\\
	f(\f)&=\le(\f-\f_H\ri)\le[f_H^\f+\sum\limits_{k\geq1}b_k^{f,\f}\le(\f-\f_H\ri)^k\ri]\;,\\
	B(\f)&=\frac{1}{2}\log\le(\frac{f_H^\f}{V'(\f_H)}\ri)+\sum\limits_{k\geq1}b_k^{B}\le(\f-\f_H\ri)^k\;,
\end{align}
\end{subequations}
with all series coefficients fixed in terms of the near-horizon modes and the chosen potential $V(\f)$. Inserting the near-horizon expansions of $\f(u)$, eq.~\eqref{phiNH}, into eqs.~\eqref{NHphi} and equating the result with the near-horizon solution of $A(u)$ and $B(u)$, eqs.~\eqref{NHu}, relates $f_H^\f$ to the near-horizon mode $f_H$ in the $u$-coordinate, ensuring that near-horizon solutions satisfy eq.~\eqref{uphiReln}. In particular, it follows from eq.~\eqref{uphiReln} and $\f(u)$'s near-horizon expansion, eq.~\eqref{phiNH}, that
\begin{align}
	f_H=-\frac{L^2\,V'(\f_H)}{2}\,\frac{e^{B(\f_H)}}{L}\;.\label{NHmodeReln}
\end{align}

In order to determine the boundary conditions that we must impose on the fields for the spacetime to be asymptotically $AdS_5$, let us switch to a radial coordinate $\z$ in terms of which the line element \eqref{uGauge} reads
\begin{align}
	\dif s^2=g_{mn}^{(0)}\dif x^m\dif x^n=e^{2A}\le[-f\,\dif t^2+\dif\underline{x}^2\ri]+\frac{L^2}{\z^2f}\dif \z^2\;. \label{zGauge2}
\end{align}
For this metric to approach $AdS_5$ as $\z\rightarrow0$, eq.~\eqref{zGauge}, $A$ and $f$ need to behave as
\begin{align}
	A\sim\log\le(\frac{L}{\z}\ri)\;,&&f\sim1\;.
\end{align}
Combining this with the leading near-boundary behaviour of the scalar,
\begin{align}
	\f\sim\Lambda\,\z\;,
\end{align}
one finds that 
\begin{align}
	A=-\log\f+\log\le(\Lambda\,L\ri)+o(1)\;,&&f=1+o(1)\;. \label{phiNB1}
\end{align}
Imposing these boundary conditions and setting the sub-leading mode $\log\le(\Lambda\,L\ri)$ of $A$ to zero by scaling the operator source $\Lambda$ to $\Lambda=1/L$, near-boundary solutions to \eqref{phiEOM} assume the form
\begin{align}
	A(\f)&=-\log\f+\mathcal{O}(\f^2)\;,\;\;\; f(\f)=1+\mathcal{O}(\f^4)\;,\;\;\;B(\f)=\log\le(\frac{L}{\f}\ri)+\mathcal{O}(\f^2)\;.\label{phiNB2}
\end{align}
Matching eq.~\eqref{phiNB1} with the near-boundary expansions of $A(u)$ and $\f(u)$, eq.~\eqref{ANB}, provides the relation\footnote{Compare with eq.~\eqref{AbphiL}.}
\begin{align}
	A_b=\le(\frac{\Lambda\,L}{\f_L}\ri)^2=\frac{1}{\f_L^2}\;. \label{connectionCheck}
\end{align}

To construct global solutions that connect the local solutions \eqref{NHphi} and \eqref{phiNB2}, we employed the method developed in ref.~\cite{Gubser:2008ny}. The key step is the realisation that a decoupled non-linear second-order equation for $G(\f)\equiv A'(\f)$ can be derived from eqs.~\eqref{phiEOM}\footnote{This possibility is hinted at by the observation that $A(\f)$'s near-horizon expansion turns out to be independent of the mode $f_H^\f$, and by the fact that $A(\f)$ enters eqs.~\eqref{phiEOM} only through its derivative.}:
\begin{align}
	\frac{G'(\f)}{G+V/\le(3V'(\f)\ri)}=\frac{\dif}{\dif \f}\log\le(\frac{G'(\f)}{G(\f)}+\frac{1}{6G(\f)}-4G(\f)-\frac{G'(\f)}{G+V/\le(3V'(\f)\ri)}\ri)\;.\label{masterEqn}
\end{align}
Imposing regularity at the horizon, eq.~\eqref{ANH}, global solutions to eq.~\eqref{masterEqn} depend on the single integration constant $\f_H$ and can readily be produced numerically. Solutions for $A(\f)$, $B(\f)$ and $f(\f)$ are then obtained by simple integrations of the equations of motion \eqref{phiEOM}. They depend on four additional integration constants which are fixed by requiring that the spacetime be asymptotically $AdS_5$, eq.~\eqref{phiNB2}, and that $f(\f)$ vanish at the horizon:
\begin{subequations}
\label{phiSoln}
\begin{align}
	A(\f)&=-\log\f+\int\limits_0^\f\dif\varphi\le(G(\varphi)+\frac{1}{\varphi}\ri)\;,\\
	B(\f)&=\log\le(\frac{L}{\f}\ri)+\int\limits_0^\f\dif\varphi\le(\frac{G'(\varphi)+1/6}{G(\varphi)}+\frac{1}{\varphi}\ri)\;,\\
	f(\f)&=\frac{\int_\f^{\f_H}\dif\varphi\,\exp\le[-4A(\varphi)+B(\varphi)\ri]}{\int_0^{\f_H}\dif\varphi\,\exp\le[-4A(\varphi)+B(\varphi)\ri]}\;.
\end{align}
\end{subequations}
These solutions depend on the single constant $\f_H$ which parameterises the single physical parameter $T/\Lambda=T\,L$.

Ultimately, we are looking for global solutions $A(u)$, $f(u)$, $\f(u)$ in terms of the $u$-coordinate, which we found a lot more convenient when dealing with perturbations of the metric. For this purpose, we first determine the expansion of a global solution \eqref{phiSoln} near the horizon $u=1$ by computing the modes $A_H=A(\f_H)$ and $f_H$ from relation \eqref{NHmodeReln}, and plugging the result into the near-horizon expansions \eqref{NHu} of $A(u)$, $f(u)$, and $\f(u)$. We obtain a global solution without further numerical integrations by matching this near-horizon expansion directly with the near-boundary expansion \eqref{NBu} at an intermediate value of $u$ where both local solutions are valid. The fact that the near-horizon modes stem from a global solution to the connection problem ensures that these solutions indeed display the appropriate near-boundary behaviour, which can be verified by checking relation \eqref{connectionCheck}.

For the computation of $f_H$ it is helpful to use
\begin{align}
	G(\f_H)=-\frac{1}{3}\frac{V(\f_H)}{V'(\f_H)}\;,&&\f\,G(\f)=\f\,A'(\f)\xrightarrow{\f\rightarrow0}-1\;,
\end{align}
which follow from eqs.~\eqref{ANH} and \eqref{phiNB2}, in order to simplify expression \eqref{NHmodeReln} as follows:
\begin{align}
	f_H&=-\frac{L^2\,V'(\f_H)}{2}\,\frac{e^{B(\f_H)}}{L}\nn\\
	&=\le(\frac{L^2\,V(\f_H)}{6G(\f_H)}\ri)\frac{1}{\f_H}\exp\le\{\lim\limits_{\f\rightarrow0}\le[\log\le(\frac{-G(\f_H)}{-G(\f)}\ri)+\log\le(\frac{\f_H}{\f}\ri)\ri]+\int\limits_0^{\f_H}\dif\varphi\frac{1}{6G(\varphi)}\ri\}\nn\\
	&=-\frac{L^2\,V(\f_H)}{6}\,\exp\le\{\int\limits_0^{\f_H}\dif\varphi\,\frac{1}{6G(\varphi)}\ri\}\;.
\end{align}
Hawking temperature $T$ and entropy density $s$, eq.~\eqref{Hawking}, are then given by
\begin{subequations}
\label{numericTD}
\begin{align}
	T\,L&=\frac{f_H\,e^{A_H}}{2\pi}=-\frac{L^2\,V(\f_H)}{12\pi}\frac{1}{\f_H}\exp\le\{\int\limits_0^{\f_H}\dif\varphi\le(G(\varphi)+\frac{1}{\varphi}+\frac{1}{6G(\varphi)}\ri)\ri\}\;,\\
	4G_Ns&=e^{3A_H}=\frac{1}{\f_H^3}\exp\le\{3\int\limits_0^{\f_H}\dif\varphi\le(G(\f_H)+\frac{1}{\varphi}\ri)\ri\}\;.
\end{align}
\end{subequations}
The leading high-temperature asymptotics of \eqref{numericTD} are obtained by taking the limit $\f_H\rightarrow0$, recalling that $V(0)=-12/L^2$,
\begin{subequations}
\begin{align}
	T\,L&\xrightarrow{\f_H\rightarrow0}\frac{1}{\pi\f_H}\exp\le\{\lim\limits_{\f_H\rightarrow0}\le(\int\limits_0^{\f_H}\dif\varphi\,G(\varphi)\ri)\ri\}\;,\\
	4G_Ns&\xrightarrow{\f_H\rightarrow0}\frac{1}{\f_H^3}\exp\le\{3\lim\limits_{\f_H\rightarrow0}\le(\int\limits_0^{\f_H}\dif\varphi\,G(\varphi)\ri)\ri\}\;.
\end{align}
\end{subequations}
Note that the limit of the remaining integral does not vanish by virtue of the fact that $G(\f)$, whose equation of motion \eqref{masterEqn} has regular singular points at $\f=0$ and $\f=\f_H$, does not behave smoothly in the limit $\f_H\rightarrow0$.\footnote{We believe that this point was overlooked in ref.~\cite{Gubser:2008ny}.} However, the limiting value can be computed by comparison with the perturbative high-temperature solution, eqs.~\eqref{deltaphiModes} and \eqref{sTpert}:
\begin{subequations}
\begin{align}
	T/\Lambda=T\,L&\xrightarrow{\f_H\rightarrow0}\frac{1}{\pi\f_H}\frac{\Gamma(3/4)^2}{\sqrt{\pi}}\;,\\
	4G_Ns&\xrightarrow{\f_H\rightarrow0}\le(\frac{\Gamma(3/4)^2}{\f_H\sqrt{\pi}}\ri)^3\;.
\end{align}
\end{subequations}

\paragraph{Details on the numerics}

For integrals \eqref{phiSoln}--\eqref{numericTD} in the $\f$-coordinate, a near-horizon expansion of $G(\f)$ to eleventh order was used in the region $\f_H-\f<10^{-2}$. We verified that the dependence of $A_H$ and $f_H$ on $\f_H^4$, which is sub-leading compared to the $\f_H^2$-contribution from appendix \ref{backreaction} but completely dictated by the quartic term \eqref{Vcondition} common to all potentials, is the same for all solutions.  The local solutions \eqref{NHu} and \eqref{NBu} in the $u$-coordinate were expanded to sixteen orders beyond horizon and boundary modes. Inverting the near-boundary expansion, we extracted the boundary modes $A_b$, $f_b$, $\f_L$, $\f_{SL}$ by matching the two local solutions at $u=0.5$. The numerical error due to the truncation of the two series is of order $0.5^{16}\sim 8\cdot 10^{-6}$. We checked that relations \eqref{Abreln} and \eqref{connectionCheck} between horizon and boundary modes were indeed satisfied with a numerical error smaller than $10^{-5}$ by all considered solutions. Directly matching the two local solutions in the $u$-coordinate, rather than numerically integrating from the horizon towards the boundary, involved the somewhat tedious inversion of the near-boundary expansion to sixteen orders. However, it greatly simplified the computation of the transport coefficients because the integrals \eqref{Yintegral} over the global background solution split into two simple integrals over the near-horizon and the near-boundary series solutions.


\section{Transport coefficients in the Chamblin-Reall background}\label{ChamblinReall}

In this appendix we present our results for hydrodynamic transport in $(3+1)$-dimensional holographic theories dual to a $(4+1)$-dimensional Chamblin-Reall backgrounds, which does not admit an asymptotic AdS region \cite{Chamblin:1999ya}. These are finite-temperature solutions to Einstein gravity coupled to a scalar with exponential potential, $V(\f)\propto e^{\g\f}$. If the the parameter $\g$ can be written as
\begin{align}
	\g^2=\frac{2(D-4)}{3(D-1)}
\end{align}
with integer $D>4$ the corresponding Chamblin-Reall background can be obtained from pure gravity in an $AdS_{D+1}$-black brane background via toroidal compactification \cite{Gubser:2008ny}. For such compactifications the AdS/CFT dictionary can essentially be borrowed from the higher dimensional non-compactified AdS \cite{Mateos:2007vn,Kanitscheider:2008kd} and the transport coefficients of the non-conformal fluid dual to Chamblin-Reall are completely determined by the coefficients of a $D$-dimensional conformal fluid dual to $AdS_{D+1}$. Einstein's equations are smooth under changes in $\g$ so that one can analytically continue the results to Chamblin-Reall backgrounds with arbitrary $\g$ \cite{Kanitscheider:2009as}.

To obtain the first- and second-order transport coefficients for the Chamblin-Reall background we first have to consider the conformal fluid dual to $AdS_{D+1}$. The constitutive relation of the stress tensor for a conformal fluid in $D$ dimensions up to second order in the derivative expansion was constructed in ref.~\cite{Baier:2007ix} generalising eq.~\eqref{constituent2}. The first- and second-order transport coefficients specific to the conformal fluid dual to the $AdS_{D+1}$-black brane solution are \cite{Haack:2008cp,Bhattacharyya:2008mz}
\begin{align}
	\label{conformalCoefficients}
	\eta &= \frac{s}{4\pi}\;, & \tau_\pi &= \frac{1}{4\pi T}\le[D+ \mathcal{H}\le(\frac{2-D}{D}\ri)\ri]\;, & \kappa &= \frac{D\,s}{8(D-2)\pi^2 T}\;,\\
	\l_1 &= \frac{D\,s}{32 \pi^2 T}\;, & \l_2 &= \frac{s}{8\pi^2 T}\,\mathcal{H}\le(\frac{2-D}{D}\ri)\;, & \l_3 &= 0\;,\nn
\end{align}
where $\mathcal{H}(\alpha)=\int_0^{1}\dif x\,(1-x^\alpha)/(1-x)$ denotes the harmonic number. The coefficients given in eq.~\eqref{conformalCoefficients} generalise previous results that were obtained in various fixed numbers of dimensions \cite{Baier:2007ix,Bhattacharyya:2008jc,Loganayagam:2008is,VanRaamsdonk:2008fp,Bhattacharyya:2008ji,Natsuume:2008iy}.

The first- and second-order transport coefficients of the non-conformal fluid dual to the Chamblin-Reall background can be determined as follows. Take the stress tensor of the $D$-dimensional conformal fluid dual to the $AdS_{D+1}$-black brane solution with known transport coefficients. Then dimensionally reduce this stress tensor on a $(D-4)$-dimensional torus. Finally extract the Chamblin-Reall transport coefficients by comparing the resulting expression with the non-conformal constitutive relation for the stress tensor, eq.~\eqref{constituent2} \cite{Kanitscheider:2009as}.

\begingroup
\allowdisplaybreaks
As a consequence we find that the coefficients of the non-conformal fluid obey the relations \cite{Romatschke:2009kr}\footnote{We believe that ref.~\cite{Romatschke:2009kr} missed the terms proportional to $\l_1$ in their equations for $\eta\tau_\pi^*$ and $\xi_2$. In fact, it was noted in ref.~\cite{Wu:2016erb} that the transport coefficients of a fluid dual to compactified D4-branes did not satisfy the equations for $\eta\tau_\pi^*$ and $\xi_2$ in the form they had been written in ref.~\cite{Romatschke:2009kr}. However, the results of ref.~\cite{Wu:2016erb} do satisfy our relations for $\eta\tau_\pi^*$ and $\xi_2$ which include terms proportional to $\l_1$.}
\begin{align}
	\label{ChamblinReallRelations}
	\z &= \frac{2\eta}{3}\le(1-3c_s^2\ri)\;, & \eta \tau_\pi^* &= \le(4\l_1-\eta\tau_\pi\ri) \le(1-3c_s^2\ri)\;,\\
	\kappa^* &= -\frac{\kappa}{2c_s^2}\le(1-3c_s^2\ri)\;, & \l_4 &= 0\;,\nn\\
	\tau_\Pi &= \tau_\pi \;, &\xi_1 &= \frac{\l_1}{3}\le(1-3c_s^2\ri)\;,\nn\\
	\xi_2 &= \rlap{$\displaystyle \frac{2}{9}\le[3c_s^2\eta\tau_\pi+2\l_1 \le(1-6c_s^2\ri)\ri]\le(1-3c_s^2\ri)\;,$}&&\nn\\
	\xi_3 &= \frac{\l_3}{3}\le(1-3c_s^2\ri)\;, & \xi_4 &= 0\;,\nn\\
	\xi_5 &= \frac{\kappa}{3}\le(1-3c_s^2\ri)\;, & \xi_6 &= \frac{\kappa}{3c_s^2}\le(1-3c_s^2\ri)\;\nn,
\end{align}
where $c_s^2=1/(D-1)$. These relations are indeed satisfied in the examples of compactified backgrounds for which first- and second-order transport has been studied \cite{Benincasa:2006ei,Wu:2016erb}.

In ref.~\cite{Bigazzi:2010ku} the leading order non-conformal corrections to first- and second-order transport coefficients in Chamblin-Reall backgrounds were computed using $\d\equiv 1-3c_s^2=3\g^2/2$ as small parameter for the perturbative series. In the perturbative expansion the Harmonic Number $\mathcal{H}$ appearing in eq.~\eqref{conformalCoefficients} takes more familiar values. However, given the conformal coefficients in eq.~\eqref{conformalCoefficients} and the relations \eqref{ChamblinReallRelations}, we can not only compute the leading order non-conformal corrections to the coefficients corresponding to the Chamblin-Reall background, but determine the fully non-perturbative transport coefficients listed in table~\ref{ChamblinReallCoefficients}. In particular, we can confirm explicitly that the relations $H=0$ and $\tilde{H}=0$, eqs.~\eqref{universalIdentity} and \eqref{identity}, are satisfied for the fluids dual to Chamblin-Reall backgrounds. This generalises the result of ref.~\cite{Bigazzi:2010ku} that $H=0$ is satisfied including the leading non-conformal corrections. Moreover, this implies that any holographic model described by Einstein gravity coupled to a scalar whose potential asymptotically approaches an exponential satisfies the relations $H=0$ and $\tilde{H}=0$ in the deep IR. In addition, the transport coefficients approach the values given in table~\ref{ChamblinReallCoefficients}. In particular, the results of this section apply to the deep IR of the family of models described by potential $V_{(2)}$, eq.~\eqref{V2}, which asymptotically approaches an exponential for large values of the scalar, $L^2\,V_{(2)}\rightarrow -e^{\g\f}/\g^4$, as noted in subsection~\ref{numFlows}.
\endgroup

\begingroup
\def\arraystretch{1.75}
\begin{table}[h]
\centering
\begin{tabular}{| c | c | c |}
\hline
	\multirow{2}{*}{Coefficient} & \multicolumn{2}{c|}{Chamblin-Reall background}\\ \cdashline{2-3}
	& Non-perturbative result & Leading non-conformal correction\\ \hline\hline
	$c_s^2$ & $\frac{1}{3}(1-\d)$ & $\frac{1}{3}(1-\d)$ \\ \hline
	$\eta$ & $\frac{s}{4\pi}$ & $\frac{s}{4\pi}$ \\ \hline
	$\zeta$ & $\frac{s}{6\pi}\d$ & $\frac{s}{6\pi}\d$ \\ \hline
	$\tau_\pi$ & $\frac{1}{4\pi T}\big[\frac{4-\d}{1-\d}+\mathcal{H}\big(\frac{2+\d}{\d-4}\big)\big]$ & $\frac{1}{2\pi T}\big[2-\log 2+\frac{3(16-\pi^2)}{32}\d+\cO\big(\d^2\big)\big]$ \\ \hline
	$\kappa$ & $\frac{s(4-\d)}{8\pi^2 T(2+\d)}$ & $\frac{s}{4\pi^2 T}\big[1-\frac{3}{4}\d+\cO\big(\d^2\big)\big]$ \\ \hline
	$\l_1$ & $\frac{s(4-\d)}{32\pi^2 T(1-\d)}$ & $\frac{s}{8\pi^2 T}\big[1+\frac{3}{4}\d+\cO\big(\d^2\big)\big]$ \\ \hline
	$\l_2$ & $\frac{s}{8\pi^2 T}\mathcal{H}\big(\frac{2+\d}{\d-4}\big)$ & $-\frac{s}{4\pi^2 T}\big[\log 2 +\frac{3\pi^2}{32}\d+\cO\big(\d^2\big)\big]$ \\ \hline
	$\l_3$ & $0$ & $0$ \\ \hline
	$\tau_\pi^*$ & $\frac{\d}{4\pi T}\big[\frac{4-\d}{1-\d}-\mathcal{H}\big(\frac{2+\d}{\d-4}\big)\big]$ & $\frac{2+\log 2}{2\pi T}\d+\cO\big(\d^2\big)$ \\ \hline
	$\kappa^*$ & $-\frac{3s(4-\d)\d}{16\pi^2 T(1-\d)(2+\d)}$ & $-\frac{3s}{8\pi^2 T}\d+\cO\big(\d^2\big)$ \\ \hline
	$\l_4$ & $0$ & $0$ \\ \hline
	$\tau_\Pi$ & $\frac{1}{4\pi T}\big[\frac{4-\d}{1-\d}+\mathcal{H}\big(\frac{2+\d}{\d-4}\big)\big]$ & $\frac{1}{2\pi T}\big[2-\log 2+\frac{3(16-\pi^2)}{32}\d+\cO\big(\d^2\big)\big]$ \\ \hline
	$\xi_1$ & $\frac{s(4-\d)\d}{96\pi^2T(1-\d)}$ & $\frac{s}{24\pi^2 T}\d+\cO\big(\d^2\big)$ \\ \hline
	$\xi_2$ & $\frac{s\,\d}{72\pi^2T}\big[\frac{(4-\d)\d}{1-\d}+(1-\d)\mathcal{H}\big(\frac{2+\d}{\d-4}\big)\big]$ & $-\frac{s\log 2}{36\pi^2 T}\d+\cO\big(\d^2\big)$ \\ \hline
	$\xi_3$ & $0$ & $0$\\ \hline
	$\xi_4$ & $0$ & $0$\\ \hline
	$\xi_5$ & $\frac{s(4-\d)\d}{24\pi^2T(2+\d)}$ & $\frac{s}{12\pi^2 T}\d+\cO\big(\d^2\big)$ \\ \hline
	$\xi_6$ & $\frac{s(4-\d)\d}{8\pi^2T(1-\d)(2+\d)}$ & $\frac{s}{4\pi^2 T}\d+\cO\big(\d^2\big)$ \\ \hline
\end{tabular}
\caption{List of first- and second order transport coefficients for the non-conformal fluid dual to Chamblin-Reall backgrounds characterised by the parameter $\d=\frac{3}{2}\g^2$. In addition to the non-perturbative result we explicitly state the leading order non-conformal correction for small $\d$ in order to allow for a comparison with ref.~\cite{Bigazzi:2010ku}. The disagreement in the expressions for $\tau_\pi^*$ and $\xi_2$ is due to the incomplete relations in ref.~\cite{Romatschke:2009kr} which were used in ref.~\cite{Bigazzi:2010ku}.}
\label{ChamblinReallCoefficients}
\end{table}
\endgroup


\clearpage
\bibliographystyle{JHEP}
\bibliography{HydroAlongRG}
\end{document}